\documentclass[reprint,prl,aps]{revtex4-1}
\usepackage{times}
\usepackage{graphicx}
\usepackage{amsmath, braket, amsfonts}
\usepackage{amssymb}
\usepackage{natbib}
\usepackage{sidecap}
\usepackage{bm, color} 
\usepackage{xcolor}
\usepackage{ragged2e}
\usepackage{lipsum}
\usepackage{wrapfig,booktabs}
\usepackage{siunitx}

\newcommand{\be}{\begin{equation}}
\newcommand{\ee}{\end{equation}}
\newcommand{\m}{moiré }

\newcommand{\bperp}{B_\perp}
\newcommand{\bpar}{B_\parallel}
\newcommand{\bpt}{B_{\text{PT}}}

\newcommand{\vtg}{$V_{\rm{tg}}$}
\newcommand{\vbg}{$V_{\rm{bg}}$}

\newcommand{\pfour}{$C^{+1}_{1/4}$ }
\newcommand{\mfour}{$C^{-1}_{1/4}$ }
\newcommand{\mthree}{$C^{-1}_{1/3}$ }

\definecolor{cadmiumgreen}{rgb}{0.0, 0.42, 0.24}

\DeclareUnicodeCharacter{03BD}{{$\nu$}}

\makeatletter
\def\maketitle{
\@author@finish
\title@column\titleblock@produce
\suppressfloats[t]}
\makeatother

\begin{document}

\title{Topological electronic crystals in twisted bilayer--trilayer graphene}

\author{Ruiheng Su$^{1,2*}$}
\author{Dacen Waters$^{3,4*}$}
\author{Boran Zhou$^{5}$}
\author{Kenji Watanabe$^{6}$}
\author{Takashi Taniguchi$^{7}$}
\author{Ya-Hui Zhang$^{5}$}
\author{Matthew Yankowitz$^{3,8\dagger}$}
\author{Joshua Folk$^{1,2\dagger}$}

\affiliation{$^{1}$Quantum Matter Institute, University of British Columbia, Vancouver, British Columbia, V6T 1Z1, Canada}
\affiliation{$^{2}$Department of Physics and Astronomy, University of British Columbia, Vancouver, British Columbia, V6T 1Z1, Canada}
\affiliation{$^{3}$Department of Physics, University of Washington, Seattle, Washington, 98195, USA}
\affiliation{$^{4}$Intelligence Community Postdoctoral Research Fellowship Program, University of Washington, Seattle, Washington, 98195, USA}
\affiliation{$^{5}$Department of Physics and Astronomy, Johns Hopkins University, Baltimore, Maryland, 21205, USA}
\affiliation{$^{6}$Research Center for Electronic and Optical Materials, National Institute for Materials Science, 1-1 Namiki, Tsukuba 305-0044, Japan}
\affiliation{$^{7}$Research Center for Materials Nanoarchitectonics, National Institute for Materials Science, 1-1 Namiki, Tsukuba 305-0044, Japan}
\affiliation{$^{8}$Department of Materials Science and Engineering, University of Washington, Seattle, Washington, 98195, USA}
\affiliation{$^{*}$These authors contributed equally to this work.}
\affiliation{$^{\dagger}$myank@uw.edu (M.Y.); jfolk@physics.ubc.ca (J.F.)}

\date{\today}

\maketitle

\textbf{In a dilute two-dimensional electron gas, Coulomb interactions can stabilize the formation of a Wigner crystal~\cite{wigner1934interaction,grimes1979wigner,Andrei1988}. Although Wigner crystals are topologically trivial, it has been predicted that electrons in a partially-filled band can break continuous translational symmetry and time-reversal symmetry spontaneously to form a form of topological electron crystal known as an anomalous Hall crystal~\cite{dong2023theory,zhou2023fractional,dong2023anomalous,kwan2023anomalous,sheng2024quantum,tan2024parent,soejima2024anomalous,dong2024stability}. Here, we report the observation of a generalized version of the anomalous Hall crystal in twisted bilayer--trilayer graphene, whose formation is driven by the \m potential. The crystal forms at a band filling factor of one electron per four moir\'e unit cells ($\nu=1/4$) and quadruples the unit-cell area, coinciding with an integer quantum anomalous Hall effect. The Chern number of the state is exceptionally tunable, and can be switched reversibly between $+1$ and $-1$ by electric and magnetic fields. Several other topological electronic crystals arise in a modest magnetic field, originating from $\nu=1/3$, $1/2$, $2/3$, and $3/2$. The quantum geometry of the folded bands is likely very different from that of the original parent band, enabling possible future discoveries of correlation-driven topological phenomena.
}

Berry curvature is the foundational ingredient required for the formation of topological electronic states. When time-reversal symmetry is broken spontaneously, these states give rise to dramatic experimental signatures at zero magnetic field. Quantum anomalous Hall (QAH) insulators are a notable example, characterized by precise quantization of the Hall conductance to integer multiples of the square of the electron charge divided by Planck's constant, $Ce^2/h$~\cite{haldane1988model,RevModPhys.95.011002}. This quantization is robust against sample variations, protected by a topological Bloch-band invariant known as the Chern number, $C$, that is equivalent to the integrated Berry curvature of the filled bands in the Brillouin zone~\cite{thouless1982quantized,berryphase2010effects}. QAH insulators have been observed in several van der Waals platforms, including graphene~\cite{sharpe2019emergent,serlin2020intrinsic,stepanov2021competing,polshyn2022topological,chen2020tunable,lu2024fractional} and transition metal dichalcogenide \m lattices~\cite{li2021quantum,cai2023signatures,zeng2023thermodynamic,park2023observation,tao2024valley}, and rhombohedral-stacked multilayer graphene~\cite{han2023large,sha2024observation}. In each of those examples, flat electronic bands are susceptible to the spontaneous breaking of valley (and therefore time-reversal) symmetry by electron--electron interactions. The finite Chern number of the filled bands then defines the topological state of the system.

\begin{figure*}[!ht]
    \centering
    \includegraphics[width=\textwidth]{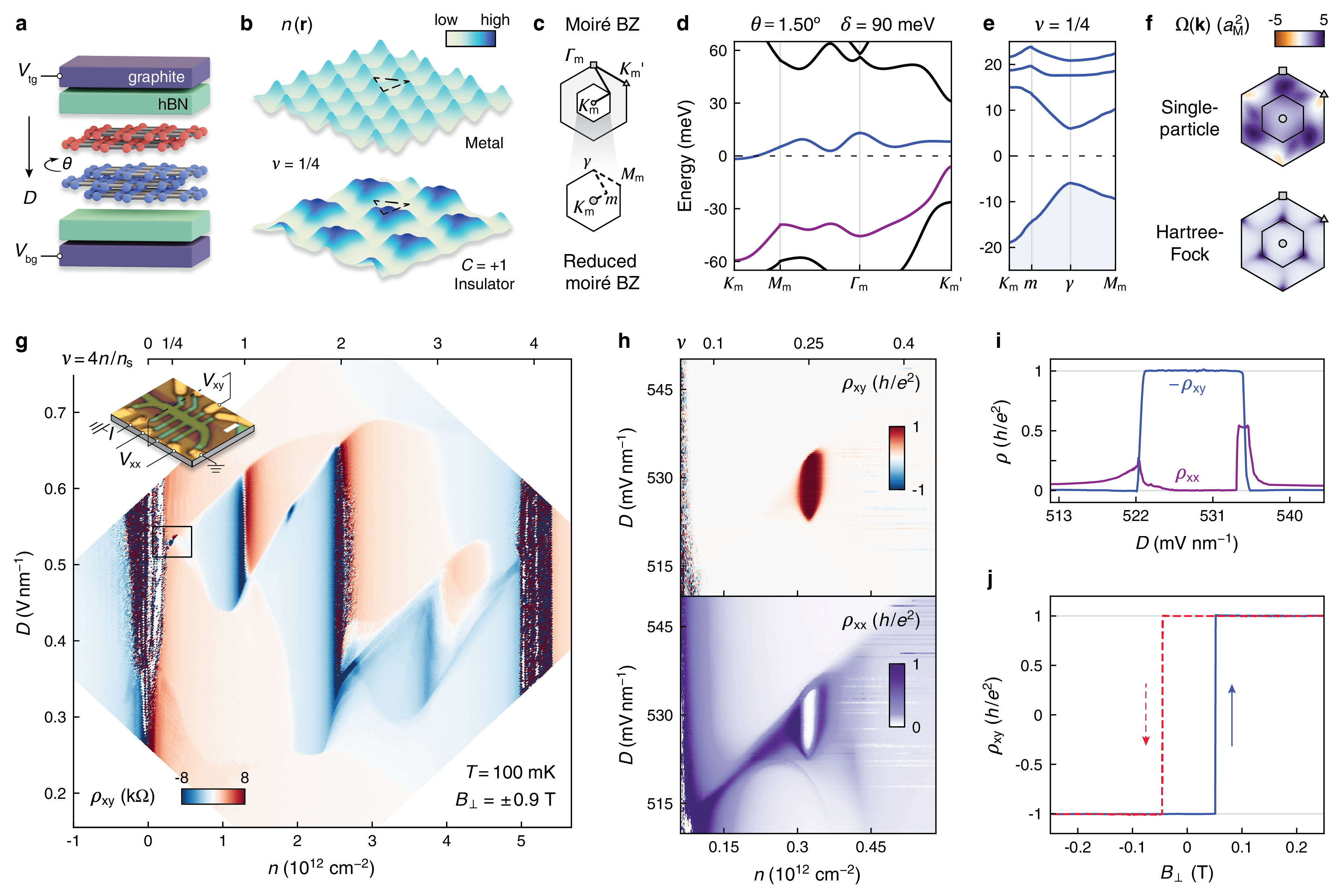} 
    \caption{\textbf{Topological electronic crystal at $\nu=1/4$ in twisted bilayer--trilayer graphene.} \textbf{a}, Schematic of the dual-gated twisted bilayer--trilayer graphene device indicated with the direction of positive electric displacement field $D$. \textbf{b}, Spatial distribution of carrier density, $n(\mathbf{r})$, at a filling of $\nu = 1/4$. The upper panel is from a single-particle calculation and corresponds to a metallic state. The lower panel is from a Hartree--Fock calculation and corresponds to a $C=1$ insulator that spontaneously quadruples the \m unit cell area. \textbf{c}, Illustration of the \m Brillouin zone (BZ) and the reduced \m BZ with the quadrupled real-space unit cell, labeled with high symmetry points. \textbf{d}, Single-particle calculated band structure plotted along the trajectory through the \m BZ shown in \textbf{c}, for $\theta=1.50^{\circ}$ and $\delta=90$~meV. \textbf{e}, Hartree--Fock band structure calculated at $\nu = 1/4$, plotted along the trajectory through the reduced \m BZ in \textbf{c}. The lowest band (shaded blue) is fully filled, with $C=+1$. Dashed lines at zero energy in \textbf{d-e} denote the Fermi energy at $\nu=1/4$.
    \textbf{f}, Single-particle (top) and Hartree--Fock (bottom) calculations of the Berry curvature, $\Omega(\mathbf{k})$, within the \m BZ as illustrated in \textbf{c}. Hartree--Fock results were divided by $4$ for clarity.
    \textbf{g}, Antisymmetrized map of $\rho_{xy}$ obtained at $T = 100$~mK and $|B_{\perp}|=0.9$~T. The top axis ($\nu$) shows the corresponding number of electrons per \m unit cell. The inset shows an optical micrograph of the device labeled with the measurement configuration. The white scale bar is 2 $\mu$m. \textbf{h}, Antisymmetrized $\rho_{xy}$ and symmetrized $\rho_{xx}$ maps obtained in the boxed region in \textbf{g} at $|B_{\perp}|=10$~mT. \textbf{i}, Measurements of $\rho_{xy}$ and $\rho_{xx}$ versus $D$ at fixed $\nu = 0.251$ and $B_{\perp}=-50$~mT. \textbf{j}, $\rho_{xy}$ measured as $\bperp$ is swept back and forth at $\nu=0.253$ and $D=532$~mV\,nm$^{-1}$, $T = 10$~mK.
    }
    \label{fig:1}
\end{figure*}

Electron--electron interactions can also induce the spontaneous breaking of translational symmetry, creating a crystal of electrons known as a Wigner crystal. Wigner crystals are topologically trivial, usually pinned by disorder in two dimensions to behave as electrical insulators. However, there are decades-old predictions that topological states resembling the Wigner crystal can also exist at high magnetic field~\cite{tevsanovic1989hall}, creating a  topological electronic crystal (TEC) known as a Hall crystal. When these crystals form at zero magnetic field due to spontaneously broken time-reversal symmetry, the result is an anomalous Hall crystal that exhibits both a charge gap and a quantized anomalous Hall effect.

Although the Wigner crystal requires ultra-strong interactions to emerge, a generalized version arises more readily in \m systems, aided by the commensuration energy of the \m lattice.  Such states correspond to an electronic crystal with an integer-multiple enlargement of the unit cell area, breaking the discrete translational symmetry of the \m lattice. This ``generalized" Wigner crystal~\cite{ung2023competing, morales2023magnetism} has been extensively studied in WSe$_2$/WS$_2$ heterobilayers \cite{regan2020mott,li2021imaging}. By analogy, an anomalous Hall crystal may also be seeded by a \m potential, giving rise to a topological electronic crystal that breaks the discrete translational symmetry of the moir\'e lattice (Figs~\ref{fig:1}a,b), folding the original \m bands into a smaller Brillouin zone (Figs~\ref{fig:1}c--e) and multiplying the number of bands by the same integer that describes the unit-cell enlargement. The Chern number, $C$, of any given folded band is ultimately determined by its total Berry curvature, which may be substantially modified from that of the parent band by the formation of the crystal (Fig.~\ref{fig:1}f).

Here, we report the observation of an array of TEC states with $|C|=1$ in twisted bilayer--trilayer graphene. One of these states exhibits a robust integer quantum anomalous Hall effect in the absence of an external magnetic field, appearing at a doping of one electron per four \m unit cells ($\nu=1/4$ out of a total of $\nu=4$ in the lowest moir\'e conduction band). This state appears to correspond to a generalized version of the AHC, with strong pinning to commensurate expansions of the moir\'e lattice. A sequence of additional TEC states form in a modest magnetic field, corresponding to $\nu=1/3$, $1/2$, $2/3$ and $3/2$. These are reminiscent of earlier observations of TEC states in various moir\'e lattices, both with~\cite{wang2015evidence,spanton2018observation,saito2021hofstadter,xie2021fractional,he2023symmetry,xie2024strong} and without~\cite{xie2021fractional,polshyn2022topological} an external magnetic field, but are distinct in several important ways. They are the first to establish a clear disconnect between the Chern number of the folded band ($C=1$ at fractional $\nu$) and that of the parent state at integer band filling ($C=0$ at $\nu=1$ and $2$), highlighting the ability to generate topological folded bands upon crystallization. Additionally, the Chern number of the state is exquisitely tunable with external fields --- the electric displacement field, $D$, and both the in-plane, $B_{\parallel}$, and out-of-plane, $B_{\perp}$, magnetic fields --- enabling unprecedented control of its topological properties. The extreme sensitivity to $B_{\parallel}$ is especially unusual for an orbital magnet.

\begin{figure*}[!ht]
    \centering
    \includegraphics[width=\textwidth]{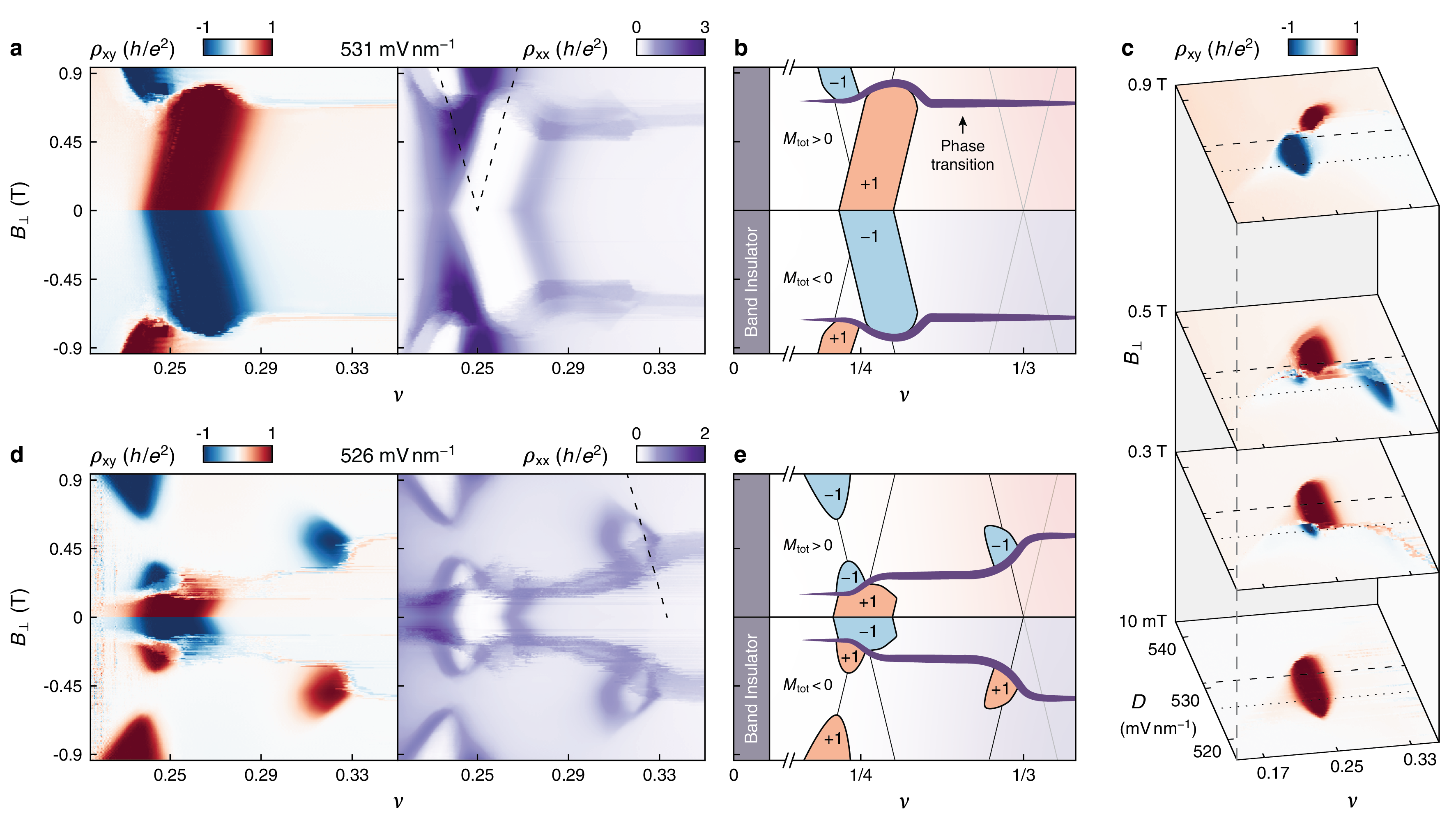}
    \caption{\textbf{Competing topological states at fractional band filling.} \textbf{a}, Landau fan of antisymmetrized $\rho_{xy}$ (left) and symmetrized $\rho_{xx}$ (right), obtained at fixed $D = 531$~mV$\,$nm$^{-1}$, $T = 10$~mK, with $\nu$ as the fast measurement axis. Dashed lines show the expected evolution of $\vert C \vert = 1$, $\nu = 1/4$ states ($C_{1/4}^{ +1 }$, $C_{1/4}^{ -1 }$) based on the Streda formula. \textbf{b}, Schematic diagram depicting the primary features in \textbf{a}. The total magnetization of the state, $M_{tot}$, is assumed to be aligned with $\bperp$. 
    \textbf{c}, Antisymmetrized $\rho_{xy}$ maps obtained at several values of $\bperp$. The dashed (dotted) line indicates $D = 531$~mV$\,$nm$^{-1}$ ($526$~mV$\,$nm$^{-1}$).
    \textbf{d}, Landau fan obtained at fixed $D = 526$~mV$\,$nm$^{-1}$. The dashed line shows the expected evolution of a $C = -1$, $\nu = 1/3$ state ($C_{1/3}^{ -1 }$). \textbf{e}, Schematic diagram depicting the primary features in \textbf{d}.
    }
    \label{fig:2}
\end{figure*}

\subsection{Evidence for a topological electronic crystal state}

Figure~\ref{fig:1}a shows the sample structure, consisting of Bernal-stacked bilayer and trilayer flakes twisted by $1.50^\circ$ with respect to each other (see Extended Data Fig.~\ref{efig:ED_nD_4K}), as well as top and bottom graphite gates that can be used to tune the carrier density $n$, and displacement field, $D$. The band structure depends both on the magnitude and sign of $D$, owing to the difference in the number of graphene layers above and below the \m interface. In this work, we focus on electronic states that emerge for $D > 0$, such that the conduction band is polarized towards the trilayer graphene~\cite{waters2024topological}. Figure~\ref{fig:1}d shows a representative single-particle band-structure calculation for the system (see Methods), including a potential difference between the top and bottom layers, $\delta$, to capture the effect of an applied $D$. The flat, isolated \m conduction band (blue curve) is conducive to spontaneous symmetry-breaking, essential for the TEC state to emerge. Indeed, this band has previously been observed to host multiple states that break spin and/or valley (together called isospin) degeneracy~\cite{waters2024topological}. Such states can be seen at $\nu=1$, $2$, and $3$ in a map of the antisymmetrized Hall resistivity, $\rho_{xy}$, shown in Fig.~\ref{fig:1}g (see Methods for additional details).

We focus first on one symmetry-broken state that appears very close to charge neutrality and around $D=0.53$~V nm$^{-1}$, within the boxed region in Fig.~\ref{fig:1}g. Figure~\ref{fig:1}h maps out $\rho_{xy}$ and $\rho_{xx}$ over this region near zero field. Remarkably, a stripe of large $\rho_{xy}$ close to $h/e^{2}\approx 25.81$ k$\Omega$ emerges from a comparably featureless background. The carrier density where the stripe appears, $n = 0.323 \times 10^{12}$ cm$^{-2}$, corresponds to adding one electron for every four \m unit cells ($\nu=1/4$). This quantized Hall resistance, concomitant with a vanishing $\rho_{xx}$, is a hallmark of an integer QAH state (Fig.~\ref{fig:1}i). Accordingly, the quantization of $\rho_{xy}$ persists to zero external magnetic field and is sharply hysteretic with a coercive field $B_c\approx 50$~mT (Fig.~\ref{fig:1}j), up to a temperature of $T\approx400$~mK (see Extended Data Figs.~\ref{efig:ED_Tdep1} and~\ref{efig:ED_Tdep_2}). The sign of $\rho_{xy}$ flips abruptly across $B_c$ to form a single square loop. The formation of a gapped topological state at a rational fractional value of $\nu$ is understood to correspond either to a fractional Chern insulator, or to a TEC state with simultaneous breaking of time-reversal and discrete translational symmetry. Only the latter scenario is consistent with the observed integer quantization of $\rho_{xy}$.

\begin{figure*}[!ht]
    \centering
    \includegraphics[width=\textwidth]{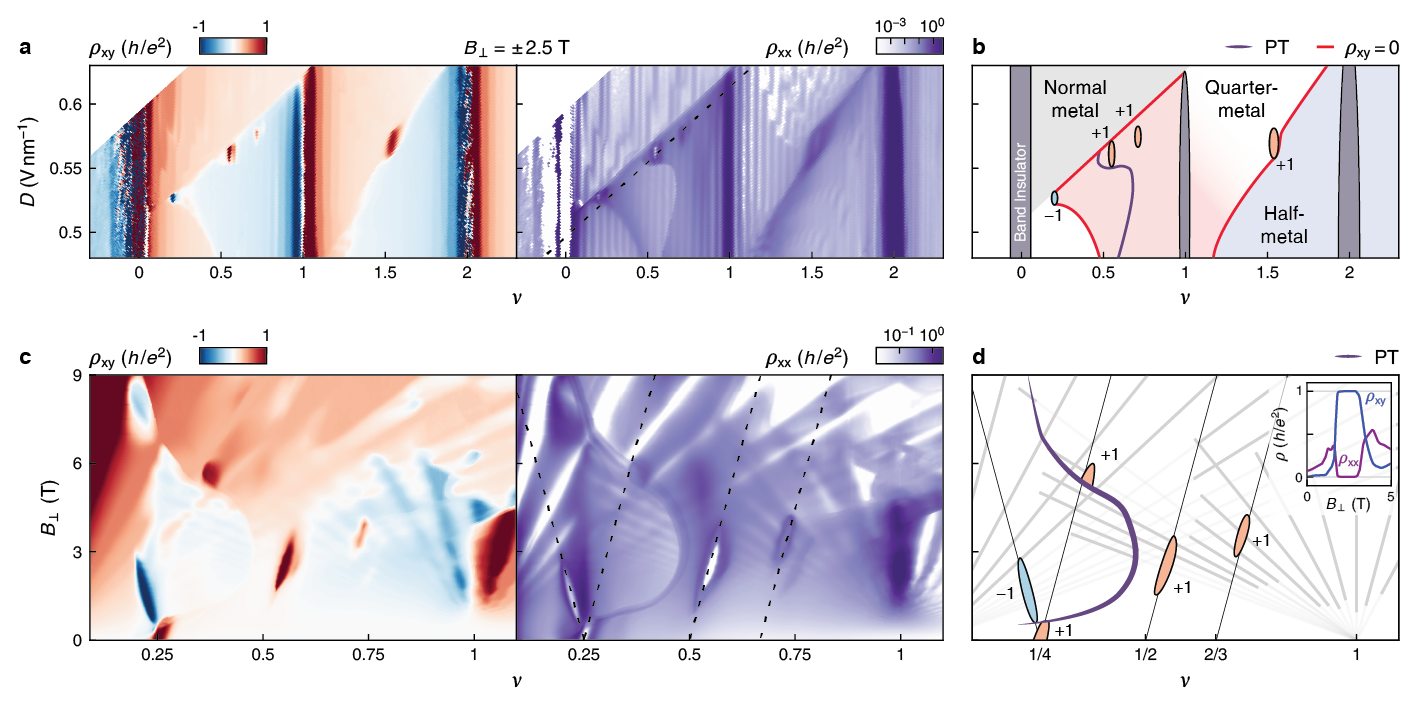}
\caption{\textbf{Topological electronic crystal states in a magnetic field.} \textbf{a}, Antisymmetrized $\rho_{xy}$ and symmetrized $\rho_{xx}$ maps obtained at $|B_{\perp}|=2.5$~T, $T = 10$~mK. \textbf{b}, Schematic phase diagram of the primary features in \textbf{a}. TEC states (blue and orange) are labeled with their respective Chern numbers. Regions of normal--, half--, and quarter--metal are assigned based on the degeneracy of low-field quantum oscillations~\cite{waters2024topological}. Trivial insulating states ($C=0$) are denoted by vertical gray stripes. Contours of $R_{xy}=0$, corresponding to either van Hove singularities or spontaneous isospin-symmetry breaking, are denoted by red curves. The purple curve denotes the same first-order phase transition (PT) shown in Figs.~\ref{fig:2}b and e. \textbf{c}, Landau fans of antisymmetrized $\rho_{xy}$ and symmetrized $\rho_{xx}$ obtained by sweeping $V_{\text{tg}}$ with fixed $V_{\text{bg}} = 4.89$~V (dashed line in right panel of \textbf{a}), $T = 10$~mK. The dashed black lines on the $\rho_{xx}$ fan show the trajectories of TEC states with $|C|=1$ tracing back to $\nu=1/4$, $1/2$, and $2/3$. \textbf{d}, Schematic diagram depicting the primary features in \textbf{c}. In addition to states that follow the conventions in \textbf{b}, gray lines indicate trajectories of quantum Hall states tracing back to $\nu=0$ and $\nu=1$, with thicker lines denoting states that are observed experimentally. The inset shows a line-cut of $\rho_{xy}$ and $\rho_{xx}$ from \textbf{c}, taken along the linear trajectory expected for a $C = 1$, $\nu = 1/2$ state.
    }
    \label{fig:3}
\end{figure*}

The quantized value $\rho_{xy}=h/e^{2}$ indicates that the state has a Chern number of $|C|=1$, which is confirmed by its evolution in $n$ with out-of-plane magnetic field (Fig.~\ref{fig:2}a). As an applied $B_{\perp}$ modifies the degeneracy of filled Chern bands, the charge density required to reach a topological gap changes with $B_{\perp}$. This is described by the Streda formula, $\frac{dn}{dB} = C\frac{e}{h}$, which relates the slope of the trajectory of a gapped state in the $(n, B)$ plane to its Chern number~\cite{streda1982quantised, streda1982theory}. The locations of the $\rho_{xy} = +h/e^{2}$ plateau and $\rho_{xx}$ minima indeed disperse to larger values of $n$ as $\bperp$ is increased from zero, in excellent agreement with the expected slope of a $C=+1$ state. 

Analogous measurements made at $\nu=1$ show that the state corresponding to a fully filled first \m band is a gapless metal at zero magnetic field, although it exhibits an anomalous Hall effect upon doping (see Extended Data Figs.~\ref{efig:ED_AHE_doping} and~\ref{efig:ED_high_field_fans}). The state transitions to an insulator with $C=0$ upon application of a small $\bperp$, with diverging $\rho_{xx}$ and an abrupt sign flip in $\rho_{xy}$, neither of which disperse in doping as $\bperp$ is raised (Extended Data Fig.~\ref{efig:ED_high_field_fans}). Thus, the $C=1$ QAH state at $\nu=1/4$ appears to emerge directly out of a quarter--metal phase associated with full isospin degeneracy lifting at $\nu=1$, as identified by the frequency of quantum oscillations formed in a small magnetic field~\cite{waters2024topological}. The disconnect between $C=1$ for the state at $\nu=1/4$ versus $C=0$ at $\nu=1$ is an identifying characteristic of the anomalous Hall crystal.

\subsection{Competing topological states}

Following the trajectory of the $\nu = 1/4$ state in Fig.~\ref{fig:2}a to increasing $B_\perp$ reveals another remarkable feature of the data: the $\rho_{xy} = +h/e^{2}$ plateau disappears above $\approx0.7$~T, simultaneous with the emergence of a $\rho_{xy}=-h/e^{2}$ plateau and $\rho_{xx}$ minimum. This state arises at a lower charge density ($\nu < 1/4$) than its zero-field counterpart, consistent with a $C=-1$ state also originating from $\nu = 1/4$. For clarity, we label these states by their Chern number and band filling upon extrapolation to $\bperp=0$: $(C=1, \nu=1/4) \equiv C^{+1}_{1/4}$ and $(C=-1, \nu=1/4) \equiv C^{-1}_{1/4}$. Apparently, the system undergoes a phase transition above a critical threshold field, $\bpt$, from \pfour into a new ground state that also has a quadrupled unit cell but the opposite sign of the Chern number (\mfour). This transition is first-order in nature, as evidenced by hysteretic jumps in resistance near $\bpt$ (see Extended Data Figs.~\ref{efig:ED_first_order}-\ref{efig:ED_phase_boundary}). The topological phases represented by the data in Fig.~\ref{fig:2}a are summarized in Fig.~\ref{fig:2}b. The sign reversal of $C$ across $\bpt$ is qualitatively distinct from the flip between $C = \pm 1$ states across the coercive field nearer to $B_\perp=0$. Whereas the latter occurs between time-reversed counterparts whose free energies either increase or decrease with $\bperp$, the former cannot be explained in the same way (see Methods). 

\begin{figure*}[!ht]
    \centering
    \includegraphics[width=\textwidth]{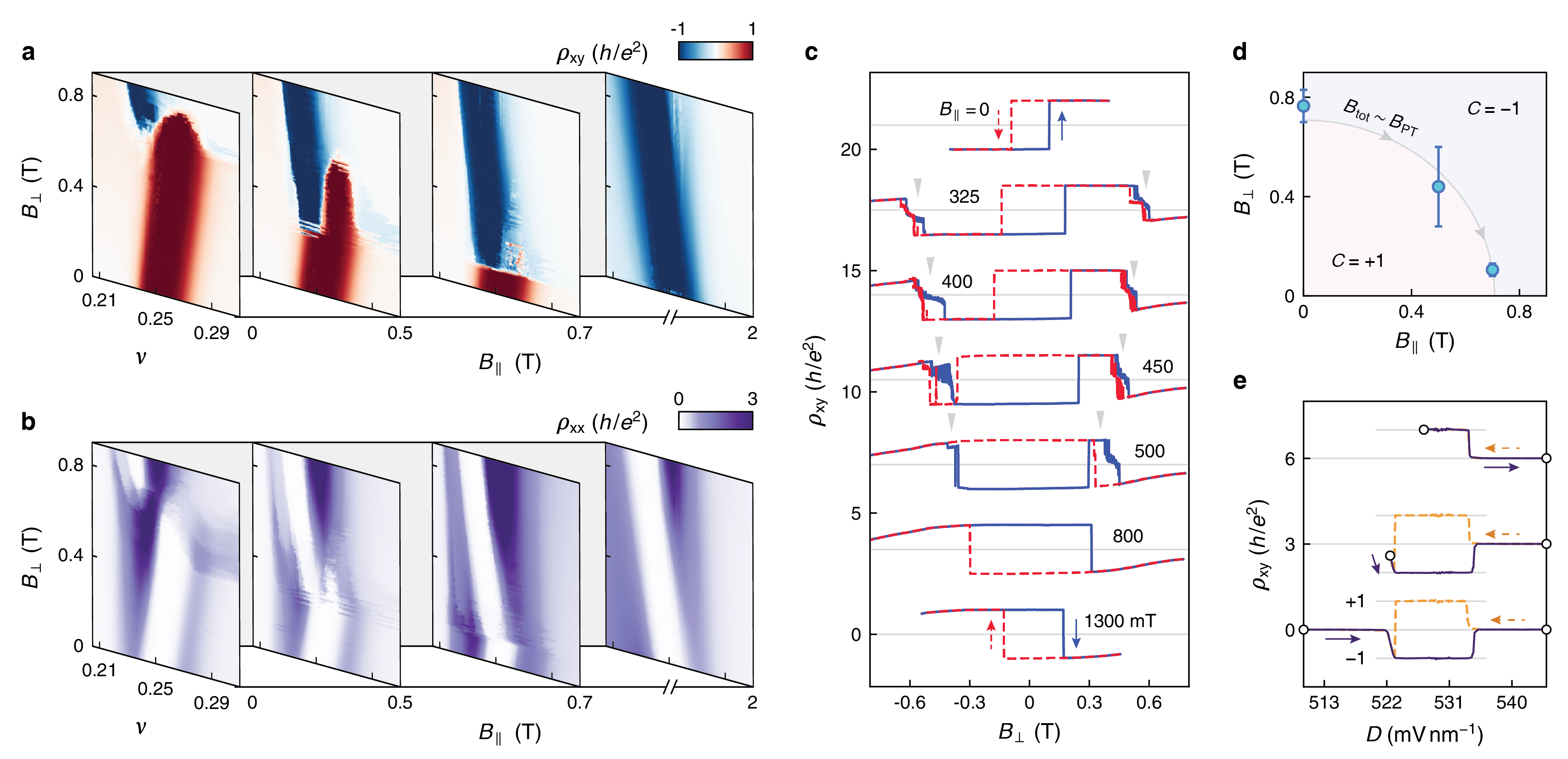}
    \caption{\textbf{Magnetic and electric field--driven transitions.} \textbf{a}, \textbf{b}, Landau fans of antisymmetrized $\rho_{xy}$ (\textbf{a}) and symmetrized $\rho_{xx}$ (\textbf{b}), collected at several values of in-plane magnetic field with $D = 531$~mV$\,$nm$^{-1}$, $T = 10$~mK. \textbf{c}, $\rho_{xy}$ measured as $\bperp$ is swept back and forth for several values of $\bpar$ ($\nu=0.253$, $D = 531$~mV$\,$nm$^{-1}$), $T = 300$~mK. \textbf{d}, Blue markers denote the values of $\bperp$ and $\bpar$ corresponding to the phase boundary separating $C=\pm 1$ states, inferred from \textbf{a} and \textbf{b}. The error bars reflect the extent to which the opposite-signed quantized plateaux overlap. The gray line traces a contour of total applied magnetic field $B_{\text{tot}}=0.71$~T. \textbf{e}, $\rho_{xy}$ measured as $D$ is swept back and forth at $\nu=0.248$, $\bperp=-0.05$~T, and $\bpar=0.35$~T. Each sweep is offset by $3h/e^2$ for clarity. The respective values $\pm h/e^2$ for each are denoted by the narrow gray lines.
    }
    \label{fig:4}
\end{figure*}

Signatures of the first-order phase transition persist across the entire range of $\nu$ shown in Fig.~\ref{fig:2}a, indicating that it is not limited to commensurate filling. The transition is also tuned by $D$, as seen in maps of $\rho_{xy}$ acquired at different fixed values of $\bperp$ (Fig.~\ref{fig:2}c). The primary features of each of these maps are the contiguous regions of red and blue coloring, which correspond to gapped states with $C=+1$ and $-1$, respectively. The Landau fans shown in Fig.~\ref{fig:2}a correspond to the value of $D$ denoted by the black dashed line. The first-order phase transition seen in Fig.~\ref{fig:2}a appears as an arc in the $(\nu,D)$ plane in Fig.~\ref{fig:2}c, mapping out the location of this phase transition in both $\nu$ and $D$. At all values of $\bperp$, the arc sharply separates regions of the maps with opposite signs of $\rho_{xy}$ in the topological gapped states, and with slightly different values of $\rho_{xy}$ in the metallic states.

\subsection{Additional magnetic-field--induced states}

The map acquired at $\bperp=0.5$~T in Fig.~\ref{fig:2}c reveals an additional pocket of quantized Hall resistance close to $\nu=0.33$, where $\rho_{xy}=-h/e^2$. Figure~\ref{fig:2}d shows a Landau fan acquired at $D=526$~mV$\,$nm$^{-1}$, corresponding to the black dotted line in Fig.~\ref{fig:2}c, which cuts through this additional blue pocket at $\bperp=0.5$~T. The sequence of phase transitions observed at this value of $D$ is more intricate, as summarized schematically in Fig.~\ref{fig:2}e. The extra blue pocket in the $\bperp=0.5$~T map from Fig.~\ref{fig:2}c can now be seen as a $\rho \sim -h/e^{2}$ plateau with concomitant $\rho_{xx}$ minimum that falls along the $C=-1$ line coming from $\nu = 1/3$ (i.e., \mthree). The appearance of the \mthree state is clearly connected with the same first-order phase transition that results in a flip between the \pfour and \mfour states. The integer Chern number and fractional filling factor of \mthree suggest a spontaneously tripled \m unit cell, contrasting with the quadrupled unit cells of \pfour and \mfour. Further raising $\bperp$ suppresses \mthree, and \mfour returns (Fig.~\ref{fig:2}d). Taken together, the data in Fig.~\ref{fig:2} indicate that the system switches sharply between a low-field state where the Chern number is $+1$ and the unit cell is quadrupled, into a higher-field state where the Chern number is $-1$ and the unit cell may be either tripled or quadrupled, depending on $\bperp$ and other parameters.

Several other TEC states arise at even higher perpendicular magnetic fields. Figure~\ref{fig:3}a shows maps of $\rho_{xy}$ and $\rho_{xx}$ at $\bperp= \pm 2.5$~T. A large portion of these maps is covered by extended vertical stripes that correspond to quantum Hall states at fixed band filling. In addition, very small pockets of suppressed $\rho_{xx}$ and enhanced $\rho_{xy}$ appear over narrow ranges of $D$, denoted by the $|C|=1$ states in the schematic diagram (Fig.~\ref{fig:3}b). With the exception of the state near $\nu=3/2$, all of these features appear along the sharp diagonal boundary that separates the isospin-unpolarized phase at large $D$ from the quarter-metal phase at smaller $D$. Notably, this phase boundary is aligned precisely to the axis corresponding to the top gate voltage, \vtg, such that the phase transition is controlled entirely by the back gate voltage, \vbg~(see Methods and Extended Data Fig.~\ref{efig:ED_gate_map} for the same maps displayed against \vtg~and \vbg). 

In order to better visualize the field-induced TEC states, we examine a Landau fan acquired along a trajectory parallel to the quarter-metal phase boundary (Fig.~\ref{fig:3}c, taken along the path denoted by the black dashed line in Fig.~\ref{fig:3}a), with salient features highlighted in Fig.~\ref{fig:3}d. There are additional states arising over small ranges of $\bperp$ that project back to $\nu=1/2$ and $2/3$. The state projecting to $\nu=1/2$ has a slope corresponding to a Chern number of $+1$ (i.e., $C^{+1}_{1/2}$), consistent with its quantized value of $\rho_{xy}=h/e^2$ (see the inset of Fig~\ref{fig:3}d). Similar behavior is seen for a more weakly developed $C=1$ state projecting to $\nu=2/3$, although the Landau fan also shows faint signs of an incipient $C=2$ state at slightly smaller $\bperp$ (i.e., $C^{+1}_{2/3}$ and $C^{+2}_{2/3}$). Another Landau fan acquired at fixed $D=0.57$~V/nm, shown in Extended Data Fig.~\ref{efig:ED_high_field_fans}a,b, also shows evidence of a weakly developed $C=1$ state projecting to $\nu=3/2$ (i.e., $C^{+1}_{3/2}$). Additionally, there are weak signatures of high-field translational symmetry breaking tracing to $\nu=3/4$, but without any associated gapped TEC states (see Landau fans in Extended Data Fig.~\ref{efig:ED_high_field_fans}c,d). We speculate that these high-field states are essentially the same as the TEC at $\nu=1/4$ but only emerge in a modest external magnetic field. 

\subsection{Chern number control of the $\nu=1/4$ state}

We now return to the nature of the sign reversal of the Chern number of the TEC at $\nu=1/4$. In addition to the first-order flip with $\bperp$, first seen in Fig.~\ref{fig:2}, we find that an in-plane magnetic field, $\bpar$, can drive a similar transition. Repeating the scans from Fig.~\ref{fig:2}a with an additional $\bpar$ applied, the transition from the $C=1$ to $C=-1$ state shifts to lower $\bperp$ for larger $\bpar$ (Fig.~\ref{fig:4}a,b), until \mfour becomes the ground state at $\bperp=0$ for sufficiently large $\bpar$ (right-most panel of Fig.~\ref{fig:4}a). This leads to a reversal in the orientation of the primary $\rho_{xy}$ hysteresis loop centered around $\bperp=0$, as seen in the high-$\bpar$ traces of Fig.~\ref{fig:4}c. For intermediate values of $\bpar$, secondary hysteresis loops are visible around $\pm\bpt$, corresponding to the flip of the Chern number in the transition between \pfour and \mfour. The primary and secondary hysteresis loops merge as $\bpar$ is raised above a crossover value. A quantitative analysis of the \pfour to \mfour crossover indicates that this transition occurs when the total magnetic field, $B_{\text{tot}}\equiv\sqrt{\bperp^2+\bpar^2}$, exceeds $\approx0.71$~T (Fig.~\ref{fig:4}d). Additionally, we find that $\bpar$ suppresses the \mthree state in the narrow ranges of $\bperp$ where it appears (Extended Data Fig.~\ref{efig:ED_effect_of_Bpar}). The sensitivity of the Chern number of the orbital magnetic state to $\bpar$ is especially unusual given its highly anisotropic nature, pointing to a potential additional role of the spin degree of freedom in determining the Chern number of the ground state.

At finite magnetic field, the reversal of the Chern state can also be driven electrically, as first indicated in Fig.~\ref{fig:2}c. The purple and yellow curves in Fig.~\ref{fig:4}e show a measurement of $\rho_{xy}$ as $D$ is swept back and forth across the region of the TEC state, sitting at $\bperp=-0.05$~T and $\bpar=0.35$~T. The hysteresis loop is now actuated by gate voltages, with quantization to $\pm h/e^2$ depending on the sweeping direction of $D$. Extended Data Fig.~\ref{efig:ED_switching} shows that this electrical switching is highly robust and repeatable, arising from the apparent lower energy of \mfour compared to \pfour at smaller values of $D$. Whereas switching of orbital magnetism with doping has been previously observed and is by now reasonably well understood~\cite{zhu2020voltage,polshyn2020electrical,grover2022chern}, the $D$-induced switching mechanism observed here is new. It implies an extreme sensitivity of the correlated electronic ground state to minute changes in the interlayer potential, and likely corresponds to distinct underlying physics.

\subsection{Discussion}

Taken together, our results reveal that TEC states driven by simultaneous time-reversal and discrete translational symmetry breaking are ubiquitous in twisted bilayer--trilayer graphene. In this context, the $\nu=1/4$ state is simply the most robust, persisting all the way down to zero magnetic field. Looking forward, scanning probe measurements to directly visualize the geometry of the electronic crystals will be critical for unraveling their precise nature. Additionally, recent theories have predicted that fractional quantum anomalous Hall (FQAH) states can emerge upon partially filling an AHC band~\cite{zhou2023fractional,dong2023anomalous}, hosting emergent quasiparticles with fractional electric charge and anyonic quantum exchange statistics. The generalized version of the AHC in twisted bilayer--trilayer graphene may provide a direct opportunity to investigate the possible relationship between topological electron crystals and the FQAH effect, since its folded bands have distinct quantum geometry from the parent band and can be doped with itinerant electrons. Finally, future experiments based on more complex device geometries --- for example, efforts aiming to proximitize the integer (or potential fractional) QAH states with a superconductor to create non-Abelian Majorana (or parafermion) modes --- will benefit enormously from the experimental simplicity of the twisted bilayer--trilayer structure, which only contains the favored Bernal stacking configuration rather than the metastable rhombohedral order. 

\section{Methods}
\textbf{Device fabrication.} Mechanically exfoliated graphene flakes with connected bilayer and trilayer regions were first identified using optical microscopy. The bilayer and trilayer regions were then separated using polymer-free anodic oxidation nanolithography~\cite{Li2018}. The van der Waals (vdW) heterostructure was sequentially assembled using a polycarbonate (PC)/polydimethylsiloxane (PDMS) stamp~\cite{wang2013one} in the following order: graphite, hBN, bilayer graphene, trilayer graphene, hBN, graphite. The graphene flakes were rotationally misaligned by $\theta\approx1.5^{\circ}$ by rotating the stage after picking up the bilayer graphene. In principle, this should create a twisted AB-ABA stacking geometry. However, we note that AB-BA stacking faults are known to exist in bilayer graphene flakes~\cite{Ju2015} and cannot be detected optically. Therefore, it is possible that the device is in a twisted BA-ABA stacking geometry, rather than the assumed AB-ABA configuration. Future work will be needed to resolve whether such a difference in stacking configuration is experimentally meaningful. 

After assembly, the completed vdW stack is dropped onto a Si/SiO$_2$ wafer. We used standard electron beam lithography and CHF$_3$/O$_2$ plasma etching to define vdW stacks into a Hall bar geometry and standard metal deposition techniques (Cr/Au)~\cite{wang2013one} to make electrical contact.

\textbf{Transport measurements.} The transport measurements were carried out across two thermal cycles, first in a Bluefors LD dilution refrigerator equipped with a 3-axis superconducting vector magnet and then in a Bluefors XLD dilution refrigerator with a one-axis superconducting magnet. In both systems, the nominal base mixing chamber temperature was $T = 10$~mK, as measured by a factory-supplied RuO$_x$ sensor. Unless otherwise specified, measurements were carried out at $T = 100$~mK. Four-terminal lock-in measurements were performed by sourcing a small alternating current of $I < 1.5$ nA at a frequency $< 20$ Hz, chosen to accurately capture sensitive transport features while minimizing electronic noise. In addition, a global bottom gate voltage of between $3.0$~V and $55.5$~V was applied to the Si substrate to improve the contact resistance.

The charge carrier density, $n$, and the out-of-plane electric displacement field, $D$, were defined according to $n= \left(C_{\text{bg}} V_{\text{bg}}+C_{\text{tg}} V_{\text{tg}}\right) / e$ and $D=\left(C_{\text{t g}} V_{\text{tg}} - C_{\text{bg}} V_{\text{bg}}\right) / 2 \epsilon_0$, where $C_{\text{tg}}$ and $C_{\text{bg}}$ are the top and bottom gate capacitance per unit area, $e$ is the elementary charge and $\epsilon_0$ is the vacuum permittivity. $C_{\rm{tg}}$ and $C_{\rm{bg}}$ were estimated using the slope of the gate-voltage--dependent Hall density, $n_{H} = 1/(e R_{H})$, where $R_{H} = \rho_{xy}/B$ is the antisymmetrized Hall coefficient. We note that in some cases, features appear to track only a single gate voltage, potentially implying a breakdown in the conversion from $(V_{bg},V_{tg})$ to $D$. Extended Data Fig.~\ref{efig:ED_gate_map} shows a prominent example of this behavior. Although we do not fully understand the origin of this effect, similar behavior is seen across a wide family of symmetry-breaking transitions in twisted Bernal graphene multilayer structures~\cite{Shen2020,Liu2020,Cao2019b,Burg2019,He2021tdbg,chen2020tMBG,polshyn2020electrical,Xu2021tmbg,waters2024topological}. We speculate that it may arise from inhomogeneous electrostatic charging of the five layers of graphene~\cite{rickhaus2019gap,kolavr2023electrostatic}. 

The carrier density, $n_{s}$, required to fully fill a four-fold degenerate \m band was estimated using the same method. Using $n_{s}$, the filling factor $\nu$ was defined according to $\nu = n/(n_{s}/4)$. A twist angle of $\theta=1.50^{\circ}$ was estimated from the doping density corresponding to adding four electrons per \m unit cell $n_{s}$ = 8$\theta^{2} / \sqrt{3} a^{2}$, where $a$ = 0.246 nm is the lattice constant of graphene.

To reduce geometric mixing between measured longitudinal and transverse voltage when $B_{\perp} > 0$, we plot the field-symmetrized values of resistance. $\rho_{xx}$ was symmetrized according to $\left(\rho_{xx}(B_{\perp} > 0) + \rho_{xx}(B_{\perp} < 0)\right)/2$, and $\rho_{xy}$ was antisymmetrized according to $\left(\rho_{xy}(B_{\perp} > 0) - \rho_{xy}(B_{\perp} < 0)\right)/2$.

The sample had multiple contacts that could be used as current or voltage probes; these are labelled in the schematic in Extended Data Fig.~\ref{efig:ED_nD_4K}, with differences highlighted in Extended Data Fig.~\ref{efig:ED_Tdep_2}. Extended Data Fig.~\ref{efig:ED_nD_4K} characterizes the sample homogeneity with maps of $\rho_{xx}$ obtained from several adjacent voltage probes. The maps are nearly identical over the accessible range of gate voltages, providing evidence for a highly homogeneous \m lattice and minimal twist angle disorder. Measurements of $\rho_{xx}$ used contacts 1 and 3, and measurements of $\rho_{xy}$ used contacts 2 and 4. However, in the following figure panels $\rho_{xx}$ and $\rho_{xy}$ measurements were based on the diagonal contact pair 1 and 4 after field symmetrization/antisymmetrization: Fig.~\ref{fig:1}h; Fig.~\ref{fig:2}a,c,d; Fig.~\ref{fig:3}a (left, $\rho_{xy}$); Fig.~\ref{fig:4}a; Extended Data Fig.~\ref{efig:ED_high_field_fans}; Extended Data Fig.~\ref{efig:ED_phase_boundary}b (left, $\rho_{xy}$). Extended Data Fig.~\ref{efig:ED_effect_of_Bpar}c used contact pair 2-3 for the $\rho_{xx}$ measurement. The source and drain contacts are (B,C) throughout.  

\textbf{Isospin symmetry breaking.}
Signatures of isospin symmetry-breaking can be easily seen in Fig.~\ref{fig:1}g. There are multiple instances in which $\rho_{xy}$ diverges and changes its sign over a narrow range of $n$ with essentially no dependence on $D$, corresponding to the chemical potential traversing an energy gap. For example, these features occur at the charge neutrality point ($n=\nu=0$), and at $n=5.17\times 10^{12}$ cm$^{-2}$ where the lowest conduction band is fully filled with four electrons per \m unit cell. Both are consistent with the expectation of single-particle gaps from the calculated band structure (Fig.~\ref{fig:1}d). Similar features at quarter-- and half--fillings ($\nu = 1,2$, and also weakly at $\nu=3$) reflect correlation-induced energy gaps between isospin-resolved bands, thus reflecting the breaking of isospin degeneracy by interactions. Our data are broadly similar to previous reports across the family of small--angle twisted multilayer graphene structures~\cite{Shen2020,Liu2020,Cao2019b,Burg2019,He2021tdbg,chen2020tMBG,polshyn2020electrical,Xu2021tmbg,waters2024topological}. 

\textbf{Comparison with previously observed TEC states.}
The state we observe at $\nu=1/4$ is the first among all moir\'e systems to exhibit a perfectly developed IQAH effect at fractional $\nu$. However, other related states have been observed in the past few years. One example is an apparent $C=+1$ state detected by compressibility measurements in a sample of magic-angle twisted bilayer graphene aligned with hBN~\cite{xie2021fractional}. The understanding of that state, however, is complicated by coexisting moir\'e potentials of similar period from the twisted graphene sheets and from the alignment of graphene and hBN. A more closely related system is twisted monolayer--bilayer graphene, which has been reported to exhibit incipient IQAH effects at $\nu=3/2$ and $7/2$. These indicate an interaction-induced doubling of the unit cell area~\cite{polshyn2022topological}, and were described as topological charge density waves. These states break the same symmetries (time-reversal and discrete moir\'e lattice translation) as the TEC we see at $\nu=1/4$, and likely share similar physical origins. Given the similarities between the twisted monolayer--bilayer and bilayer--trilayer systems from a single-particle band structure perspective~\cite{waters2024topological}, it is interesting to note that the observed states in twisted bilayer--trilayer graphene feature additional topological tunability with magnetic and electric fields, along with a new sequence of several other TEC states that arise in a modest magnetic field.

Next we turn to a comparison between the states observed in this experiment and those previously reported in rhombohedral pentalayer graphene aligned with hBN~\cite{lu2024fractional}. The AHC has been proposed as an explanation for the IQAH insulator seen in that system at $\nu=1$~\cite{zhou2023fractional,dong2023anomalous,tan2024parent,soejima2024anomalous,dong2024stability}. However, this state does not need to break any translation symmetry beyond that of the hBN--graphene \m lattice, making it challenging to distinguish from a standard Chern insulator. Although it may indeed correspond to a TEC, the importance of the moir\'e potential for its formation remains unclear. Future work will be needed to understand how TEC formation and resultant properties depend on the strength of the moir\'e potential, which is presumed to be weak in pentalayer graphene but stronger in twisted bilayer--trilayer graphene. In particular, an interesting open question is whether the AHC and its generalization to stronger superlattice potentials successfully captures the physics of TEC states within a single framework.

\textbf{Nature of the first-order phase transition.}
The nature of the first-order phase transition separating the $C=+1$ and $-1$ states at $\nu=1/4$ remains mysterious (Figs.~\ref{fig:2}-\ref{fig:3}). It appears first at $\bperp=0.7$~T, identified in the data by a sharp jump in $\rho_{xx}$ and an inversion in $\rho_{xy}$. The phase transition shifts rapidly towards larger values of $\nu$ as the field is raised and then eventually reverses direction back towards $\nu=0$. Notably, the \pfour state briefly recurs above $\bperp\approx6$~T, when the phase transition crosses back across the value of $\nu$ corresponding to the projected trajectory of the \pfour state.

One possibility is that this phase transition represents a recondensation of carriers from one valley of the graphene Brillouin zone into the other (i.e., between the two time-reversed copies of a given state). Such a scenario has been previously suggested to explain the charge-doping--induced reversal of the sign of $C$ observed in orbital magnetic states at integer $\nu$ in magic-angle twisted bilayer graphene and twisted monolayer--bilayer graphene~\cite{zhu2020voltage,polshyn2020electrical,grover2022chern}. This effect is understood to result from a flip in the sign of the total magnetization of the state upon doping across the topological gap, reversing the valley corresponding to the ground state. However, the simplest version of this scenario is not consistent with our data. First, we do not find a first-order reversal in $C$ upon doping across the $\nu=1/4$ state, except for within the small range of $\bperp$ in which there is also a field--induced reversal of the Chern number. More generally, a cursory overview of the data in Fig.~\ref{fig:2} indicates that the $C=1$ and $-1$ phases across the finite-field transition are not a time-reversed pair. In contrast with the near-perfect (anti\nobreakdash-\hspace{0pt})symmetry of the transition across $B=0$, the states above and below the $\bperp=0.1-0.3$ T transition are clearly different, with \mfour much narrower than \pfour across the boundary, and no $C^{+1}_{1/3}$ at the equivalent location to \mthree despite the fact that the transition passes diagonally through \mthree. 

An alternative possibility is that the two states represent qualitatively different electronic orders, whether within the same valley or in opposite ones. There are many degrees of freedom that could vary between these two orders, including both isospin textures and the real-space geometry of the electronic crystal. Given the sensitivity of the first-order phase transition to $\bpar$, which typically does not couple strongly to orbital degrees of freedom in 2D materials, it is possible that spin physics might play a role in distinguishing the two states. Extended Data Fig.~\ref{efig:ED_diagram} shows schematic energy diagrams of a possible two-state model consistent with our observations, agnostic to the specific ordering of each state but with the constraint that the two states have opposite Chern numbers for a given sign of their total magnetization in order to explain the flip between \mfour and \pfour across $\bpt$. 

\textbf{Band structure calculation.}
We use the Bistritzer-MacDonald continuum model to calculate the band structure. The Hamiltonian for each valley and spin is:
\begin{equation}
H=
\begin{pmatrix}
    H_{AB}&H_M^\dagger\\
    H_M & H_{ABA}
\end{pmatrix},
\end{equation}
where $H_{AB}$ and $H_{ABA}$ are the Hamiltonian of the bilayer and the trilayer respectively. $H_M$ represents the moir\'e tunneling between the bilayer and the trilayer. The bilayer Hamiltonian $H_{AB}$ is:
\begin{eqnarray}
H_{AB}= \begin{pmatrix}
            -\frac{\delta}{2} & \gamma_0\pi & \gamma_4\pi& \gamma_3\pi^\dagger \\
            \gamma_0\pi^\dagger & -\frac{\delta}{2} & \gamma_1 & \gamma_4\pi \\
            \gamma_4\pi^\dagger & \gamma_1 & -\frac{\delta}{4} & \gamma_0\pi \\
            \gamma_3\pi & \gamma_4\pi^\dagger & \gamma_0\pi & -\frac{\delta}{4}
        \end{pmatrix},
\end{eqnarray}
and the trilayer Hamiltonian $H_{ABA}$ is:
\begin{eqnarray}
&H_{ABA}=
\begin{pmatrix}
            0 & \gamma_0\pi & \gamma_4\pi& \gamma_3\pi^\dagger & \frac{\gamma_2}{2} & 0 \\
            \gamma_0\pi^\dagger & 0 & \gamma_1 & \gamma_4\pi  &0 & \frac{\gamma_5}{2}\\
            \gamma_4\pi^\dagger & \gamma_1 & \frac{\delta}{4} & \gamma_0\pi  & \gamma_4\pi^\dagger & \gamma_1\\
            \gamma_3\pi & \gamma_4\pi^\dagger & \gamma_0\pi^\dagger & \frac{\delta}{4}  & \gamma_3\pi & \gamma_4\pi^\dagger \\
            \frac{\gamma_2}{2} & 0 &  \gamma_4\pi & \gamma_3\pi^\dagger & \frac{\delta}{2} & \gamma_0\pi \\
            0 & \frac{\gamma_5}{2} &  \gamma_1 & \gamma_4\pi & \gamma_0\pi^\dagger & \frac{\delta}{2}
        \end{pmatrix},&
\end{eqnarray}
where $\pi=(k_x-\mathrm{i}k_y)e^{\pm\mathrm{i}\frac{\theta}{2}}$, with $\theta$ being the twist angle. The $\pm$ sign corresponds to the bilayer/trilayer respectively. $\delta$ is the potential difference between the top and bottom graphene layers. The hopping parameters are $(\gamma_0,\gamma_1,\gamma_2,\gamma_3,\gamma_4,\gamma_5)=(2610\frac{\sqrt{3}}{2},361,-20,-283\frac{\sqrt{3}}{2},-140\frac{\sqrt{3}}{2},20)~\mathrm{meV}$. The moir\'e tunneling is:
\begin{eqnarray}
    H_M=t_M
    \begin{pmatrix}
        \alpha & e^{-\mathrm{i}\frac{2\pi j}{3}} \\
        e^{\mathrm{i}\frac{2\pi j}{3}} & \alpha
    \end{pmatrix},
\end{eqnarray}
where $j=0,1,2$ represents the tunneling between the bottom layer of the bilayer and the top layer of the trilayer, with momentum differences given by $\mathbf{Q_j}$. The vectors $\mathbf{Q_0}=(0,0)^T, \mathbf{Q_1}=(-\frac{2\pi}{\sqrt{3}a_M},-\frac{2\pi}{a_M})^T, \mathbf{Q_2}=(\frac{2\pi}{\sqrt{3}a_M},-\frac{2\pi}{a_M})^T$. We use $t_M=110~\mathrm{meV}$ and $\alpha=0.3$ in our calculation. 

\textbf{Hartree--Fock calculations.}
We performed the Hartree--Fock calculation in the reduced Brillouin zone at $\nu=1/4$. The Coulomb interaction term is given by:
\begin{eqnarray}
    H_\mathrm{int}=\frac{1}{2A}\sum_\mathbf{q}\sum_{l,l^\prime}V_{l;l^\prime}(\mathbf{q}):\rho(\mathbf{q})\rho(-\mathbf{q}):,
\end{eqnarray}
where $\rho_l(\mathbf{q})$ is the density operator at layer $l$, and $A$ is the area of system. The interaction potential is defined as:
\begin{equation}
    V_{l,l^\prime}(\mathbf{q})=\frac{e^2e^{-q|l-l^\prime|d_{\mathrm{layer}}}\tanh{(q\lambda)}}{2\epsilon\epsilon_0|q|},
\end{equation}
 where $d_{\mathrm{layer}}=0.34~\mathrm{nm}$ is the distance between adjacent layers, $\epsilon=6$ is the dielectric constant and $\lambda=30~\mathrm{nm}$ is the screening length. We include 4 conduction bands in the reduced BZ at single-particle level and perform the self-consistent calculation. We test $100$ random initial ansatzes and select the one with the lowest energy.

\textbf{Spatial charge distribution calculation.}
The spatial distribution $n(\mathbf{r})$ in Fig.~\ref{fig:1}(b) is defined as $\langle c^\dagger(\mathbf{r})c(\mathbf{r)\rangle}$, where $c(\mathbf{r})$ is electron annihilation operator in real-space. It can be calculated as:
\begin{eqnarray}
    n(\mathbf{r})=\frac{1}{N_1N_2}\sum_\mathbf{k}\sum_{\mathbf{q}}\langle u(\mathbf{k+q})\lvert u({\mathbf{k}})\rangle e^{-\mathrm{i}\mathbf{q\cdot r}},
\end{eqnarray}
where $\mathbf{k}$ is defined in the BZ, which is discretized into $N_1\times N_2$ points. $\lvert u(\mathbf{k+q})\rangle$ is the periodic part of the Bloch wavefunction. $\mathbf{q}=n_1 \mathbf{G_1}+n_2\mathbf{G_2}$ is any integer linear combination of reciprocal lattice vectors.  The unit of $n(\mathbf{r})$ is the area of the BZ divided by $4\pi^2$. For the metallic state, we use the single-particle Bloch wavefunction and sum over $\mathbf{k}$ below the Fermi energy at $\nu=1/4$. The reciprocal lattice vectors are moir\'e reciprocal lattice vectors. For the TEC state, we use the Hartree--Fock Bloch wavefunction and sum over $\mathbf{k}$ in the reduced \m BZ. The first conduction band is fully filled at $\nu=1/4$. The reciprocal lattice vectors are half the moir\'e reciprocal lattice vectors.

\textbf{Berry curvature and Chern number calculation.}
We use the method from Ref.~\onlinecite{fukui2005chern} to calculate Berry curvature. The BZ is discretized into an $N_1\times N_2$ grid. For each $\mathbf{k}$ in the BZ, we calculate the $U(1)$ link:
\begin{eqnarray}
    U_\mu(\mathbf{k})=\frac{\langle u(\mathbf{k}+\delta k_\mu \hat{k}_\mu)\lvert u(\mathbf{k})\rangle}{N_\mu(\mathbf{k})},
\end{eqnarray}
where $N_\mu(\mathbf{k})=|\langle u(\mathbf{k}+\delta k_\mu \hat{k}_\mu)\lvert u(\mathbf{k})\rangle|$ and $\mu=1,2$. The Berry curvature is then given by:
\begin{eqnarray}
    \Omega(\mathbf{k})=-\frac{\mathrm{i}\ln W(\mathbf{k})}{\delta k_1\delta k_2},
\end{eqnarray}
where $W(\mathbf{k})$ is $U_1(\mathbf{k}) U_2(\mathbf{k}+\delta k_1 \hat{k}_1) U^{-1}_1(\mathbf{k}+\delta k_2 \hat{k}_2) U_2^{-1}(\mathbf{k})$. We ensure that $-\pi<-\mathrm{i}\ln W(\mathbf{k})\le\pi$. For the metallic state, the BZ is the moir\'e BZ. For the TEC state at $\nu=1/4$, the BZ is the reduced \m BZ. The Chern number is calculated as:
\begin{eqnarray}
    C=\frac{1}{2\pi}\int_{\mathrm{BZ}}d^2\mathbf{k} \Omega(\mathbf{k}).
\end{eqnarray}

Extended Data Fig.~\ref{efig:ED_Chern_Phase_diagram} shows the calculated Chern number of the TEC state at $\nu=1/4$ as various parameters ($\theta$, $\delta$, $\alpha$, and $\epsilon$) are tuned. Our Hartree--Fock calculations predict a $C=1$ state for realistic parameters, consistent with our experiment. However, we caution that we do not know the precise values of any of these parameters experimentally except for $\theta$, and the calculations also find other Chern number ground states within relatively narrow parameter ranges. Furthermore, Hartree--Fock predicts a TEC state over a substantially wider range of $\delta$ than is observed experimentally; this is consistent with the tendency of Hartree--Fock calculations to underestimate the formation energy of symmetry-broken states. Despite these uncertainties, the $C=1$ states found experimentally and theoretically motivate a plausible connection between the two. The sensitivity of the Chern number in Hartree--Fock calculations to the dielectric screening, $\epsilon$, highlights the importance of interactions in determining the topological properties of the electronic crystal.

The valley Chern number of the single-particle \m conduction band is calculated to be $+2$. Based on this, a spin-valley--polarized state at $\nu=-1$ is expected to have $|C|=2$ since it corresponds to a single filled valley. Because the state at $\nu=1$ in our device is metallic, it is not possible to extract its Chern number. However, the vanishing AHE at $\nu=1$ (Extended Data Fig.~\ref{efig:ED_AHE_doping}) is inconsistent with a gapless but valley-imbalanced $|C|=2$ state. Our observation of a $C=0$ correlated insulator at finite field (Extended Data Fig.~\ref{efig:ED_high_field_fans}) is also inconsistent with the prediction from the single-particle band structure calculation, although the Chern number of a correlated state at finite $\bperp$ need not be the same as its value at $\bperp=0$. Future work will be necessary to identify the origin of this apparent inconsistency. In any case, the Chern number of $1$ for the $\nu=1/4$ state is not equal to one-quarter of either of the possible Chern numbers of the parent state at $\nu=1$ (i.e., $0$ or $2$). Thus, the ambiguity in identifying the latter does not affect our primary conclusion that a combination of band folding and interactions determine the Chern number of the TEC state.

\section{Acknowledgements}
The authors thank Liang Fu, Leon Balents, Marcel Franz, Michael Zaletel, and Andrea Young for helpful discussions. Experiments at UBC were undertaken with support from the Natural Sciences and Engineering Research Council of Canada; the Canada Foundation for Innovation; the Canadian Institute for Advanced Research; the Max Planck-UBC-UTokyo Centre for Quantum Materials and the Canada First Research Excellence Fund, Quantum Materials and Future Technologies Program; and the European Research Council (ERC) under the European Union’s Horizon 2020 research and innovation program, Grant Agreement No. 951541. Work at UW was supported by National Science Foundation (NSF) CAREER award no. DMR-2041972 and NSF MRSEC 2308979. The development of twisted graphene samples was partially supported by the Department of Energy, Basic Energy Science Programs under award DE-SC0023062. D.W. was supported by an appointment to the Intelligence Community Postdoctoral Research Fellowship Program at University of Washington administered by Oak Ridge Institute for Science and Education through an interagency agreement between the US Department of Energy and the Office of the Director of National Intelligence. M.Y. acknowledges support from the State of Washington-funded Clean Energy Institute. K.W. and T.T. acknowledge support from the JSPS KAKENHI (Grant Numbers 21H05233 and 23H02052) and World Premier International Research Center Initiative (WPI), MEXT, Japan. Y.-H.Z. was supported by the National Science Foundation under Grant No. DMR-2237031. This work made use of shared fabrication facilities at UW provided by NSF MRSEC 2308979.

\section{Author Contributions}
R.S. performed the measurements in the Folk lab at UBC, and analyzed the data; D.W. made the sample and performed initial transport measurements in the Yankowitz lab at UW; M.Y. and J.F. supervised the measurements; R.S., M.Y. and J.F. wrote the manuscript with B.Z. and Y.Z. providing theory support; K.W. and T.T. provided the hBN crystals.

\section*{Competing interests}
The authors declare no competing interests.

\section*{Additional Information}
Correspondence and requests for materials should be addressed to Matthew Yankowitz or Joshua Folk.

\section*{Data Availability}
Source data are available for this paper. All other data that support the findings of this study are available from the corresponding author upon request.

\bibliographystyle{naturemag}
\bibliography{main}
\newpage
\clearpage

\onecolumngrid

\renewcommand{\figurename}{Extended Data Fig.}
\renewcommand{\thesubsection}{S\arabic{subsection}}
\setcounter{secnumdepth}{2}
\setcounter{figure}{0} 
\setcounter{equation}{0}

\onecolumngrid

\section*{Extended Data}

\begin{figure}[ht!]
    \centering
    \includegraphics[width = \textwidth]{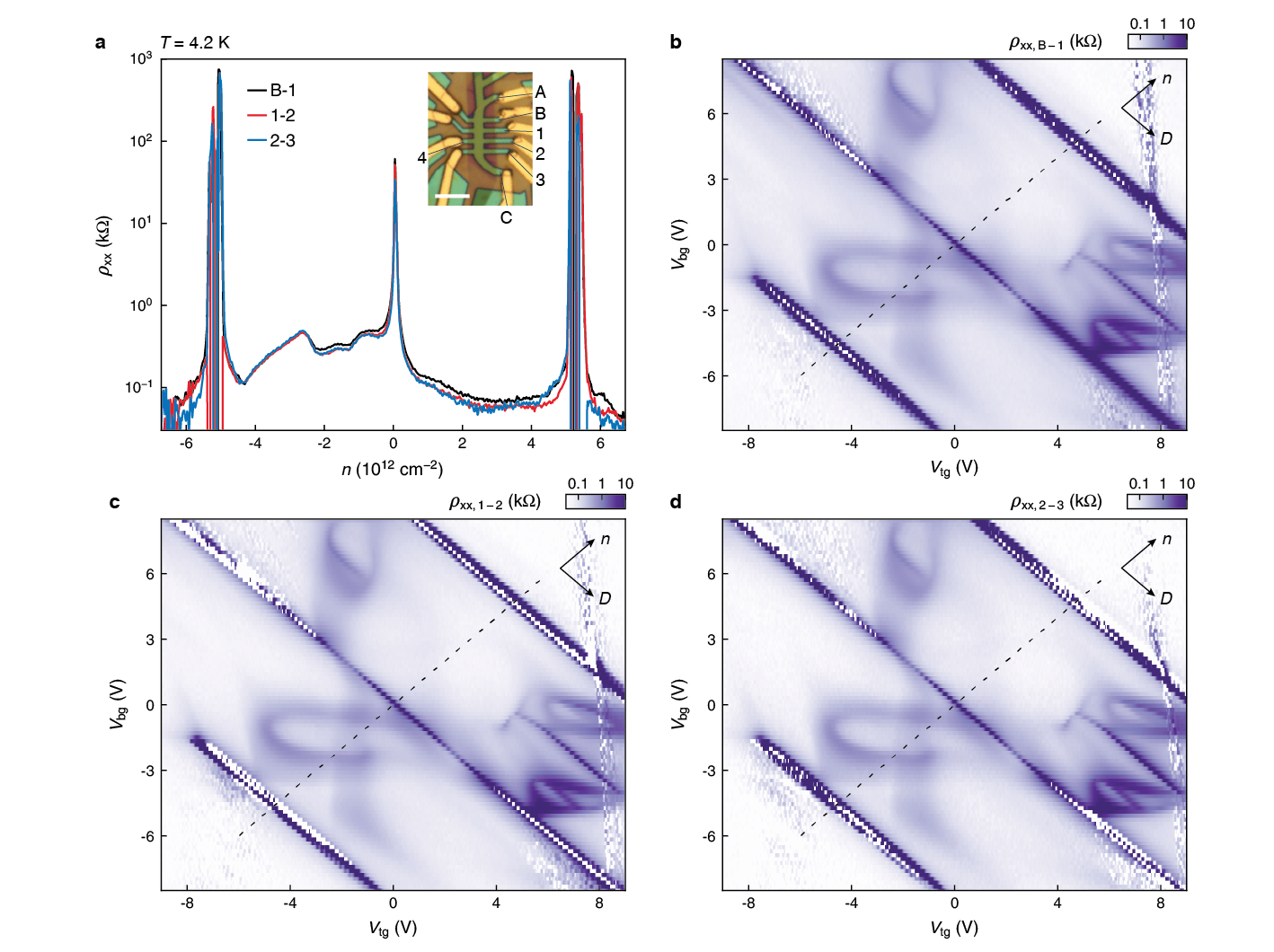}
    \caption{ \textbf{Characterization of twist angle homogeneity.} \textbf{a}, Line-cut of $\rho_{xx}(n)$ for three adjacent sets of voltage probes along the Hall bar, at fixed $D = 0$ and $T = 4.2$ K. Inset: optical micrograph of the device. Electrical contacts are labeled A-C and 1-4. The scale bar is 5 $\mu$m. \textbf{b}-\textbf{d}, Gate voltage dependence of $\rho_{xx}$ measured from contact pairs B-1, 1-2, and 2-3, respectively. The dashed line shows the contour of $D = 0$, where line-cuts in \textbf{a} were obtained. The arrows show orthogonal axes of increasing $n$ and $D$, related to \vtg~and \vbg~via the linear transformation described in the Methods.  Positive $D$ corresponds to electrons pushed to toward the bilayer, so the fully-filled valence band lives more toward the bilayer and the conduction band lives more toward the trilayer. The close alignment of $\rho_{xx}$ features for different contact pairs in \textbf{a} demonstrates the high degree of twist angle homogeneity across the length of the sample. The source and drain contacts used for \textbf{b} were A and C. All other measurements used B and C as source and drain contacts.}
    \label{efig:ED_nD_4K}
\end{figure}

\newpage
\clearpage
\begin{figure}[ht!]
    \centering
    \includegraphics[width = \textwidth]{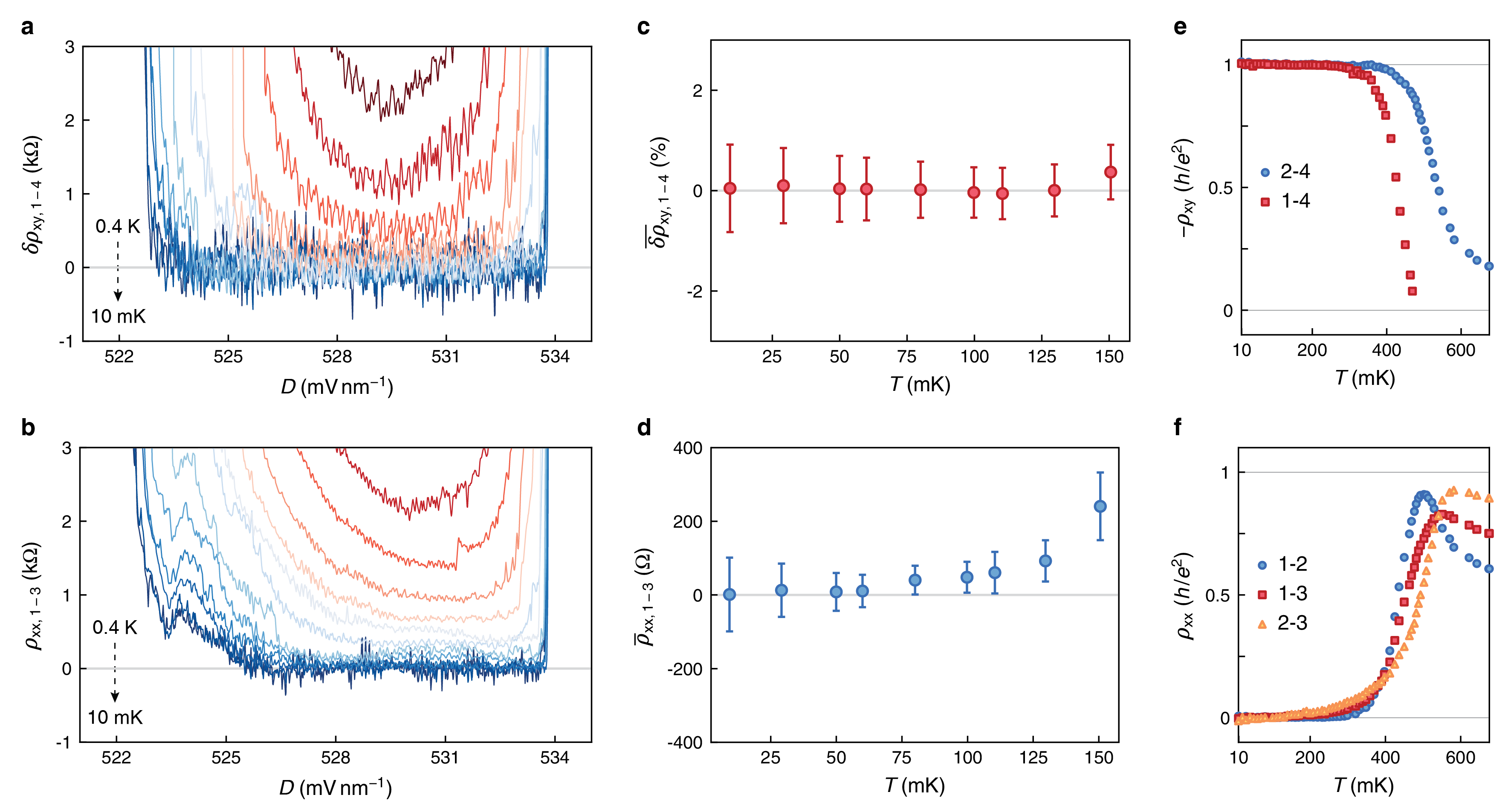}
    \caption{ \textbf{Temperature dependence of the $\nu = 1/4$ state.} \textbf{a}, Deviation of the Hall plateau from $h/e^{2}$ quantization, as a function of displacement field and temperature, obtained with a small $B_{\perp} = -50$~mT. Since $\rho_{xy} < 0$, $\delta \rho_{xy}$ is shown as $h/e^{2} + \rho_{xy}$ for clarity. Numerical subscripts (1-4 in this case) indicate the contact pairs used for the measurement, as labeled in the inset of Extended Data Fig.~\ref{efig:ED_nD_4K}a. \textbf{b}, $\rho_{xx}$ measured in the same way as \textbf{a}. \textbf{c}, Percent deviation of the Hall resistivity, $\overline{\delta \rho_{xy}} = 100 \times \delta \rho_{xy} \times (h/e^{2})^{-1}$. \textbf{d}, Longitudinal resistivity averaged using data points between $D = 526$ and $533$~mV\,nm$^{-1}$, $\overline{\rho_{xx}}$. Error bars represent one standard deviation. \textbf{e}, \textbf{f}, $- \rho_{xy}(T)$ and $\rho_{xx}(T)$, respectively, at fixed $D = 532$~mV\,nm$^{-1}$ and $B_{\perp} = -50$ mT.}
    \label{efig:ED_Tdep1}
\end{figure}

\newpage
\clearpage
\begin{figure}[h!]
    \centering
    \includegraphics[width = \textwidth]{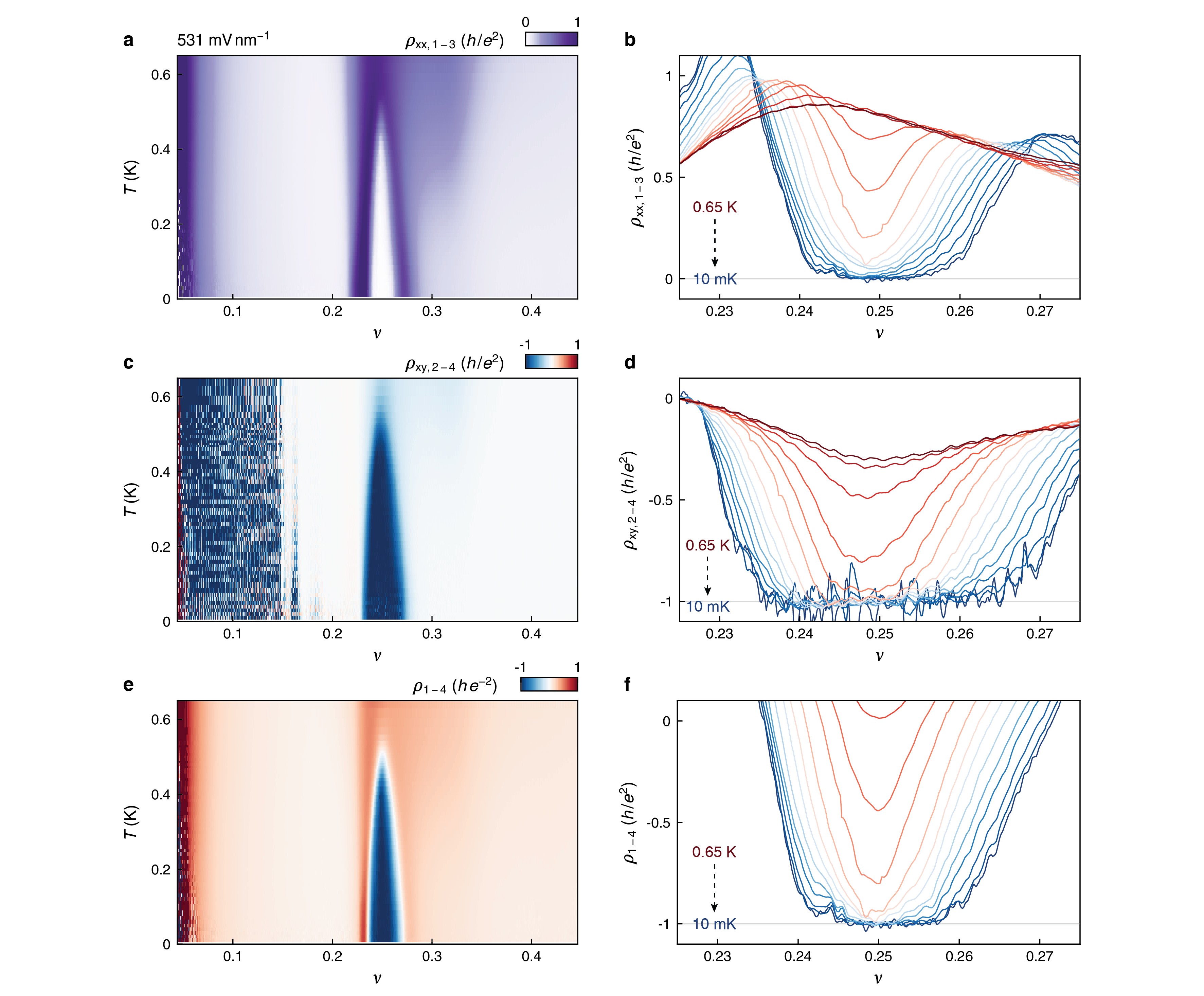}
    \caption{ \textbf{Filling factor and temperature dependence of the $\nu = 1/4$ TEC state.} \textbf{a}, Temperature dependence of $\rho_{xx, 1-3}$ as $\nu$ is tuned across the TEC state, at fixed $D = 531$~mV~nm$^{-1}$ with a small magnetic field of $B_{\perp} = -50$ mT applied. \textbf{b}, Line-cuts from \textbf{a} at selected values of $T$ between 0.65~K and 10~mK. 
    \textbf{c}-\textbf{f}, Measurements obtained under the same conditions as \textbf{a}-\textbf{b} but for \textbf{c-d} $\rho_{xy, 2-4}$ and \textbf{e-f} $\rho_{1-4}$ (corresponding to a mixture of longitudinal and Hall geometries since contacts 1 and 4 are separated by $\approx1$~$\mu$m along the length of the channel and located by opposite sides of the Hall bar). Contact 2 fails severely for $\nu\lessapprox0.15$, leading to extremely large and rapidly oscillating resistance. Similar contact issues in the TEC state at $\nu=1/4$ likely also explain the large fluctuating resistance of $\rho_{xy, 2-4}$ near base temperature. $\rho_{1-4}$ exhibits a mixture of $\rho_{xx}$-like and $\rho_{xy}$-like properties when probing metallic states but faithfully reflects the quantization of the anomalous Hall plateau in the topological gap ($\nu\approx0.25$) since the $\rho_{xx}$ contribution vanishes.       
    }
    \label{efig:ED_Tdep_2}
\end{figure}

\newpage
\clearpage
\begin{figure}[ht!]
    \centering
    \includegraphics[width = \textwidth]{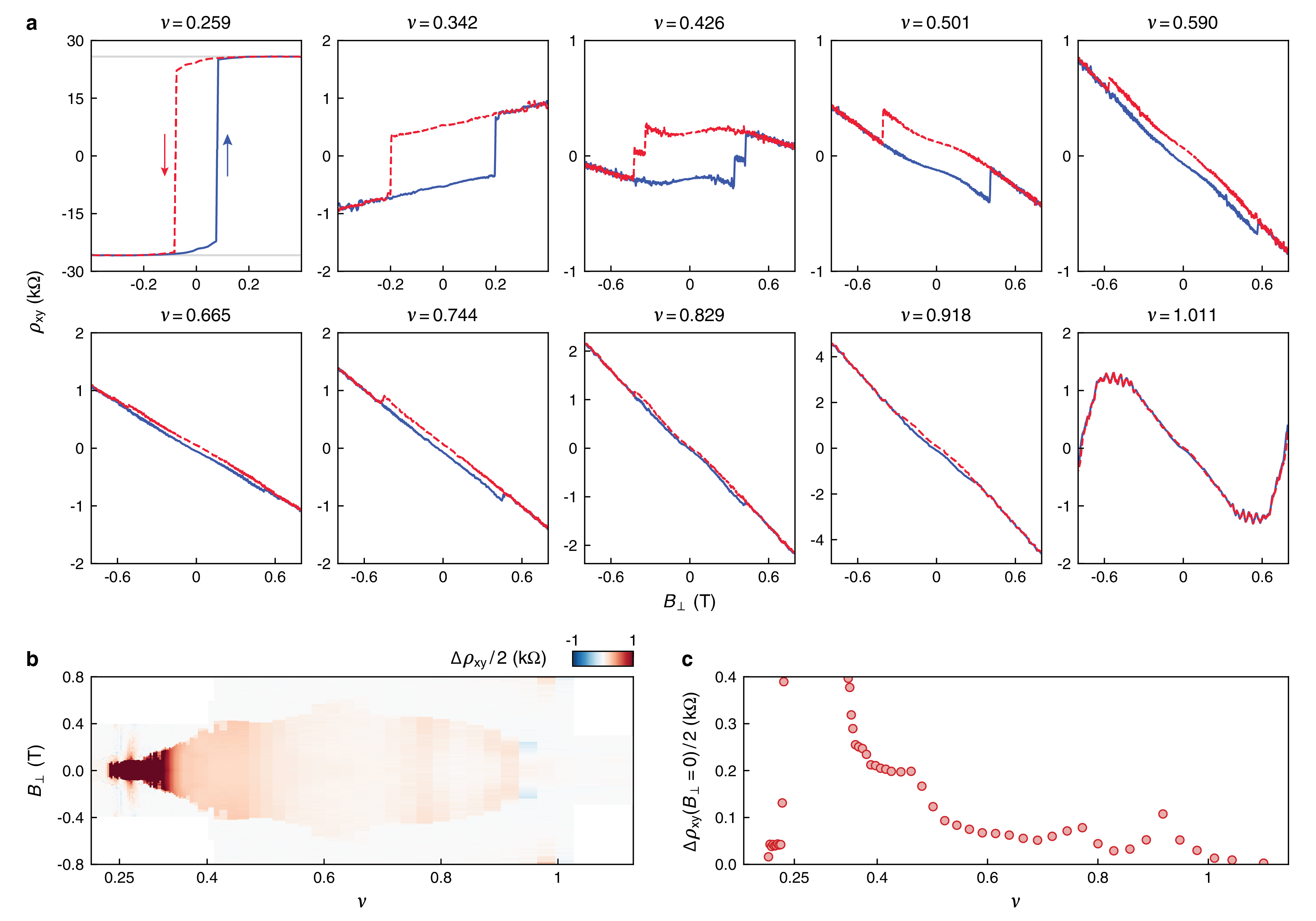}
    \caption{ \textbf{Doping dependence of the anomalous Hall effect.} \textbf{a}, Antisymmetrized $\rho_{xy}(B_{\perp})$ hysteresis loops at increasing values of $\nu$ away from $\nu = 1/4$. Data obtained at fixed $D = 532$~mV\,nm$^{-1}$ and $T = 300$ mK. The solid/dashed traces correspond to forward ($\uparrow$) and reverse ($\downarrow$) scans of $B_{\perp}$, respectively. \textbf{b}, Map of $\Delta \rho_{xy}/2 = (\rho_{xy}(B_{\uparrow}) - \rho_{xy}(B_{\downarrow}))/2$, showing the AHE over a broad range of $\nu$. The color scale is saturated to show the weakest AHE features near $\nu=1$. \textbf{c}, Line-cut of $\Delta \rho_{xy}/2$ at $B_{\perp} = 0$, showing that the small anomalous Hall effect vanishes near $\nu = 1$. }
    \label{efig:ED_AHE_doping}
\end{figure}

\newpage
\clearpage
\begin{figure}[ht!]
    \centering
    \includegraphics[width = \textwidth]{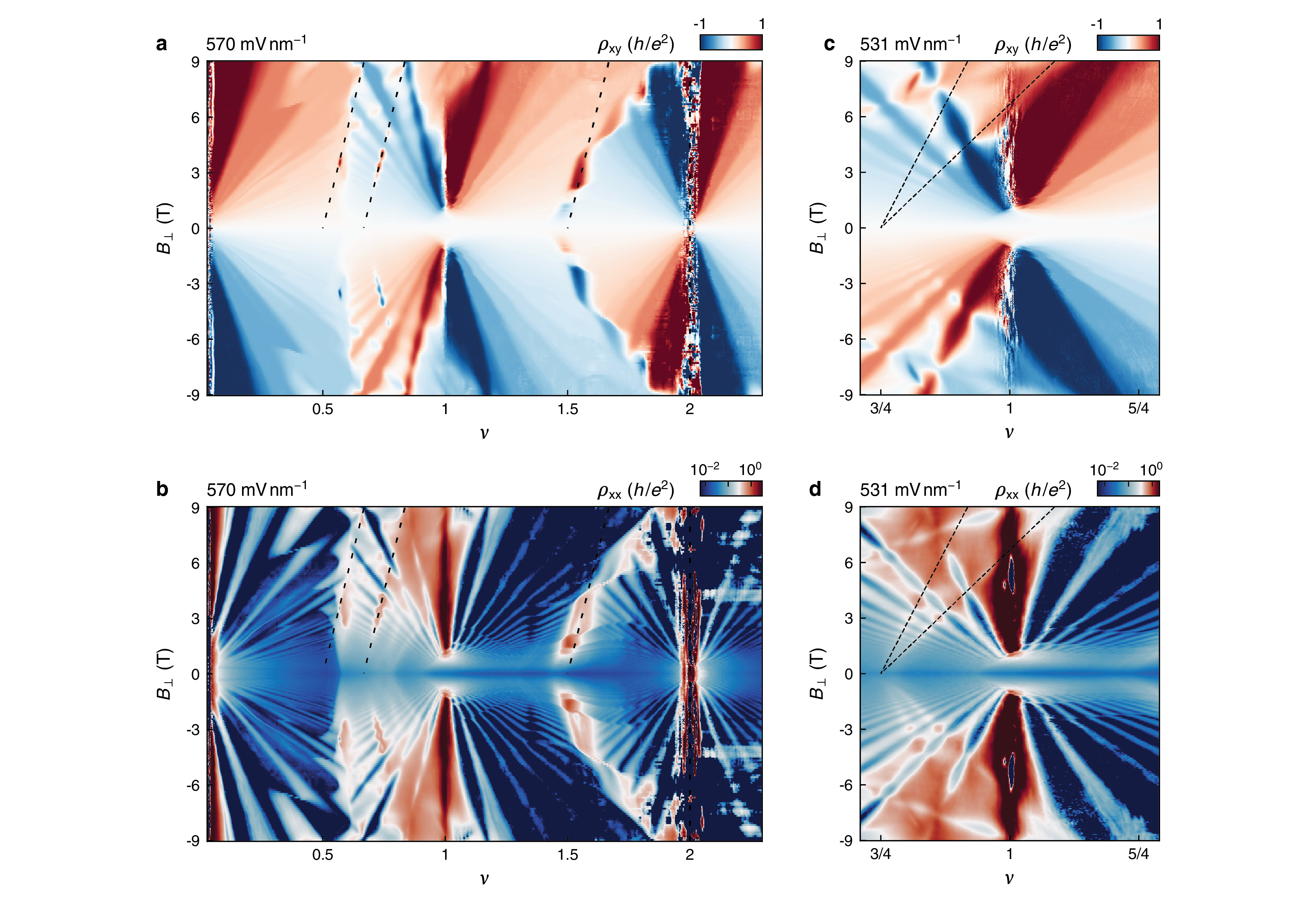}
    \caption{ \textbf{Additional Landau fans at fixed $D$.} \textbf{a}, \textbf{b}, Landau fans of antisymmetrized $\rho_{xy}$ and symmetrized $\rho_{xx}$, respectively, obtained at $D = 570$~mV\,nm$^{-1}$. Dashed lines correspond to the trajectories of states with Chern number $C = +1$ as described by the Streda formula, tracing back to $\nu = 1/2$, $2/3$ and $3/2$ at $B_{\perp} = 0$. Deviations in both $\rho_{xy}$ and $\rho_{xx}$ along these trajectories indicate the formation of TEC states. Additionally, both electron- and hole-like quantum oscillations emerge from the correlated insulators at $\nu = 1$ and $\nu = 2$ and develop into quantum Hall states at high field. At fixed $\nu = 1$ and $2$, $\rho_{xx}$ is diverging whereas $\rho_{xy} = 0$ and abruptly changes sign with $\nu$. Together, these features indicate that the correlated insulating states have $C=0$. Notably, the $\nu=1$ state only becomes insulating in a small $\bperp$ and is metallic at $\bperp\approx0$. This metal-to-insulator transition with $\bperp$ is seen irrespective of the precise value of $D$. \textbf{c}, \textbf{d}, Similar Landau fans obtained at $D = 531$~mV\,nm$^{-1}$. Dashed lines correspond to the trajectories of states with Chern numbers $C = +1$ and $+2$ tracing back to $\nu = 3/4$ at $B_{\perp} = 0$. There are weak deviations in both $\rho_{xy}$ and $\rho_{xx}$ along these trajectories, suggesting the possibility of gapless TEC states with a four-fold enlargement of the \m unit cell.
    }
    \label{efig:ED_high_field_fans}
\end{figure}

\newpage
\clearpage
\begin{figure}[ht!]
    \centering
    \includegraphics[width = \textwidth]{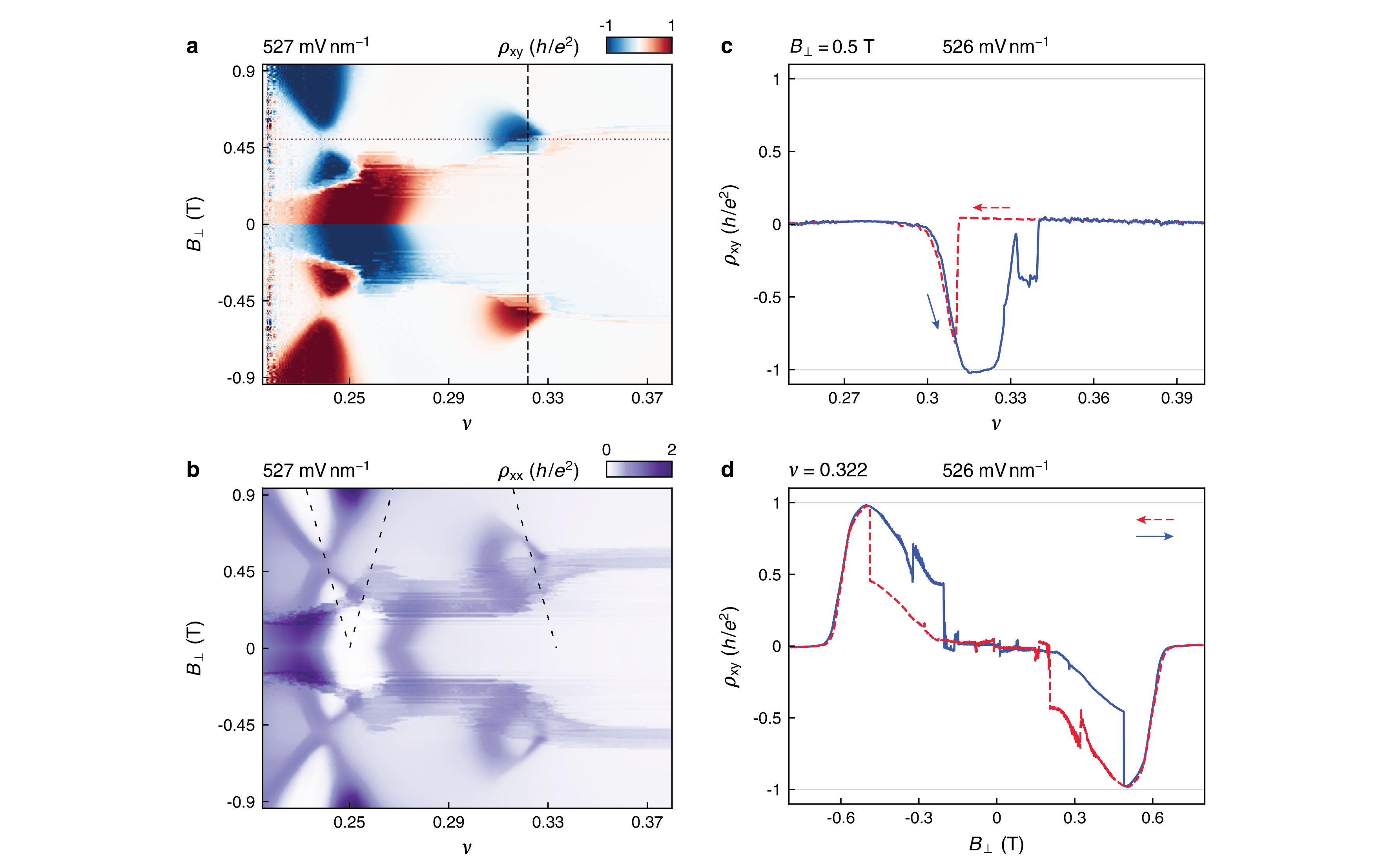}
    \caption{ \textbf{First-order phase transition and additional characterization of the $\nu=1/4$ and $1/3$ states.} 
    \textbf{a}, \textbf{b}, Landau fans of antisymmetrized $\rho_{xy}$ and symmetrized $\rho_{xx}$ obtained at fixed $D = 527$~mV\,nm$^{-1}$, respectively. (Which is larger than the $D=526$~mV$\,$nm$^{-1}$ fans shown in Fig.~\ref{fig:2}d of the main text.) Panel \textbf{a} is overlaid with dashed/dotted lines, indicating $\nu = 0.322$ and $B_{\perp} = 0.5$~T. Panel \textbf{b} is overlaid with dashed lines that show the expected trajectories of $\vert C \vert = 1$ states originating from $\nu = 1/4$ and $\nu = 1/3$. \textbf{c}, $\rho_{xy}$ for forward and reverse scans of $\nu$, at fixed $B_{\perp} = 0.5$~T and $D = 526$~mV\,nm$^{-1}$, forming a hysteresis loop as $\nu$ is swept past the phase boundary. \textbf{d}, Antisymmetrized $\rho_{xy}$ measured as $\bperp$ is swept back and forth at fixed $\nu = 0.322$, $D = 526$ mV$\,$nm$^{-1}$. There is hysteresis when the sample enters/exits the TEC state, consistent with a first-order phase transition. 
    }
    \label{efig:ED_first_order}
\end{figure}

\newpage
\clearpage
\begin{figure*}[!ht]
    \centering\includegraphics{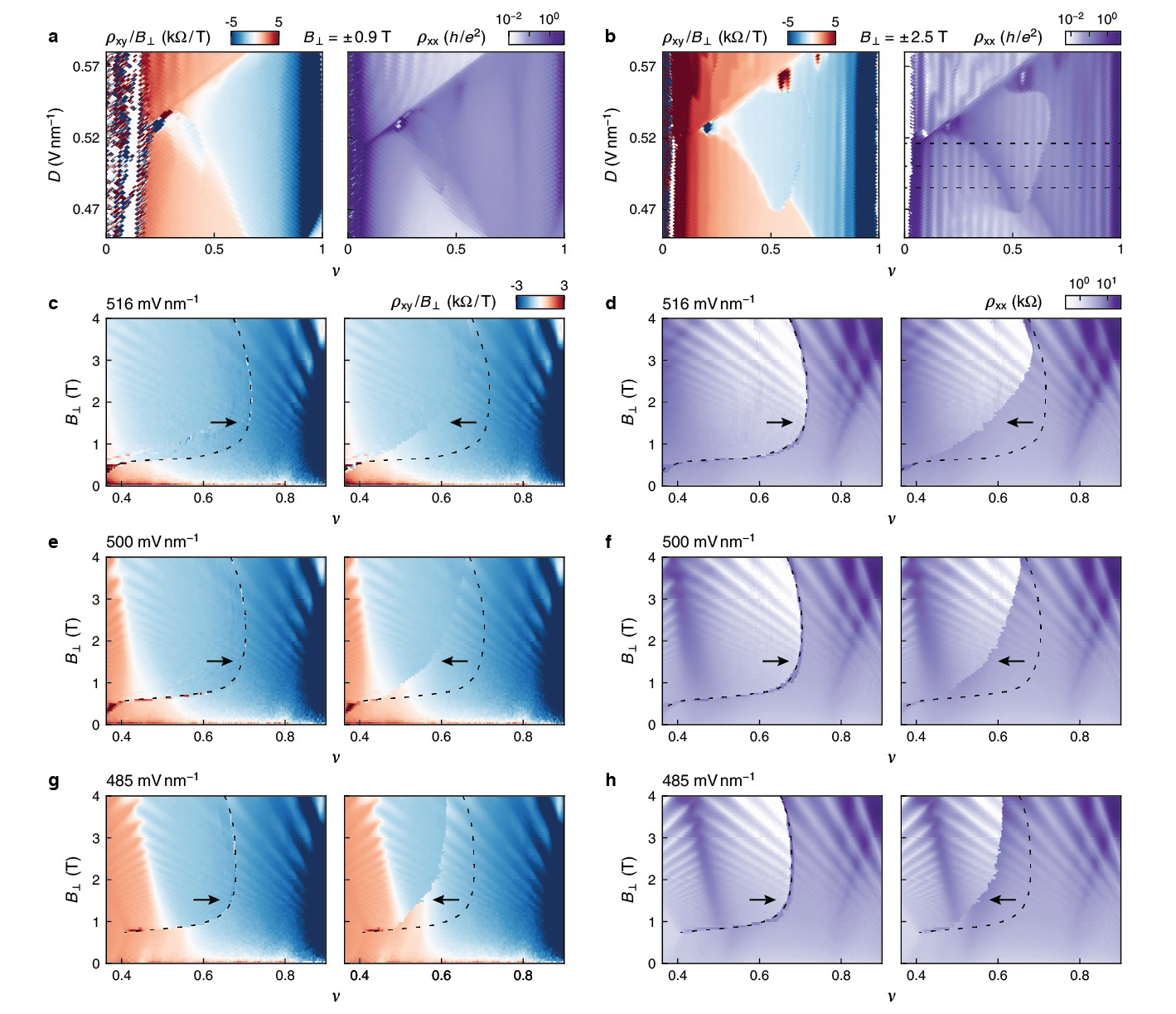}
    \caption{\textbf{Doping and magnetic field dependence of the first-order phase transition.}
    \textbf{a}, \textbf{b}, Antisymmetrized $\rho_{xy}(\nu, D)$ normalized by the applied $B_{\perp}$, and symmetrized $\rho_{xx}(\nu, D)$ at $B_{\perp} = \pm 0.9$ and $2.5$ T, respectively. \textbf{b} is reproduced from Fig.\ref{fig:3}\textbf{a}. Horizontal dashed lines mark $D = 516$, $500$, and $485$~mV\,nm$^{-1}$. \textbf{c}, Landau fan of $\rho_{xy}$ normalized by the applied $B_{\perp}$ recorded by scanning $\nu$ from away (left panel, $\rightarrow$) and towards (right panel, $\leftarrow$) charge neutrality point at each $B_{\perp}$ and $D = 516$~mV\,nm$^{-1}$. Dashed lines are a guide to the eye for the location of the phase boundary inferred from the data when $\nu$ is scanned away from the charge neutrality point. \textbf{d}, Landau fan of $\rho_{xx}$ measured at the same time as \textbf{c}. \textbf{e}, \textbf{f}, and \textbf{g}, \textbf{h}, are the same as \textbf{c},\textbf{d} but obtained at $D = 500$ and $485$~mV\,nm$^{-1}$, respectively. Both $\rho_{xx}$ and $\rho_{xy}$ exhibit hysteresis when varying $\nu$ induces a transition between the competing phases. 
    }
\label{efig:ED_phase_boundary}
\end{figure*}

\newpage
\clearpage
\begin{figure}[ht!]
    \centering
    \includegraphics[width = \textwidth]{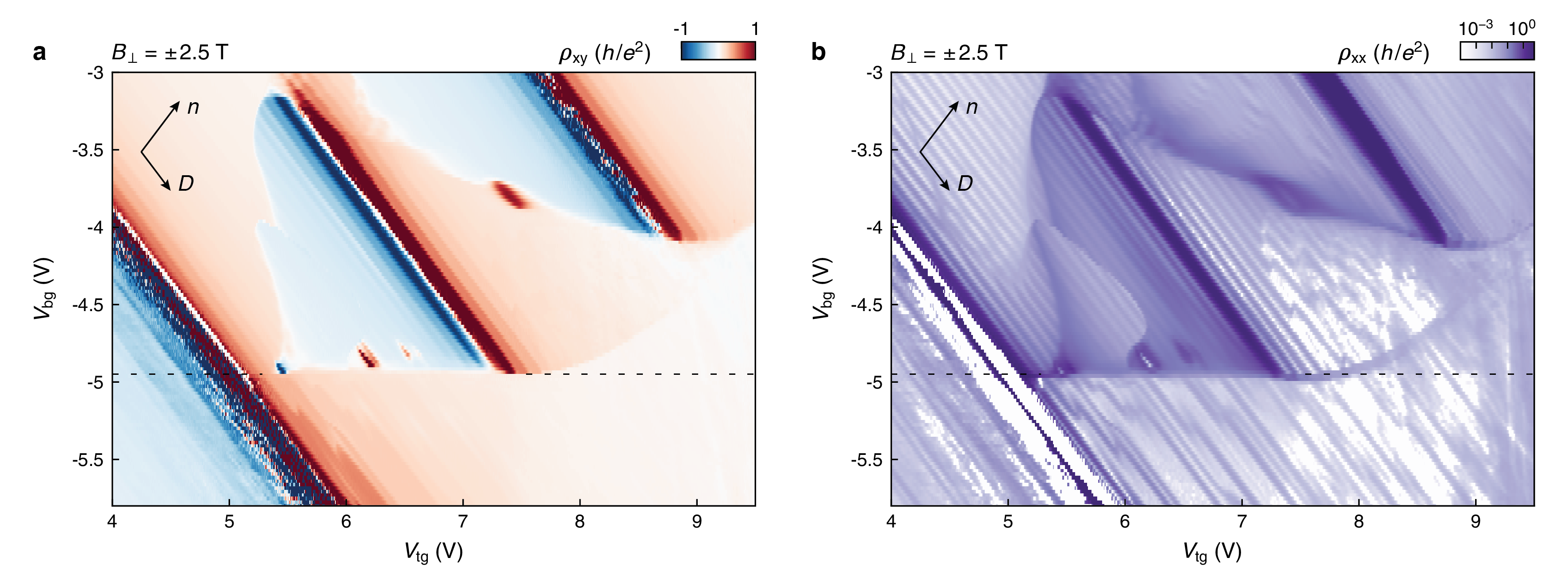}
    \caption{ \textbf{Gate--dependent resistance maps.} \textbf{a}, \textbf{b}, Antisymmetrized $\rho_{xy}$ and symmetrized $\rho_{xx}$ maps, respectively, obtained at $B_{\perp} = 2.5$~T and plotted against \vtg~and \vbg. The horizontal dashed lines mark fixed \vbg~$=4.95$~V. The arrows show orthogonal axes of increasing $n$ and $D$, related to \vtg~and \vbg~via the linear transformation described in the Methods. The dashed lines separate a region of normal metal with no broken isospin degeneracies at large negative \vbg~from a region of full isospin degeneracy lifting (sometimes coexisting with translational symmetry breaking) at smaller values of \vbg.}
    \label{efig:ED_gate_map}
\end{figure}

\newpage
\clearpage
\begin{figure}[ht!]
    \centering
    \includegraphics[width = \textwidth]{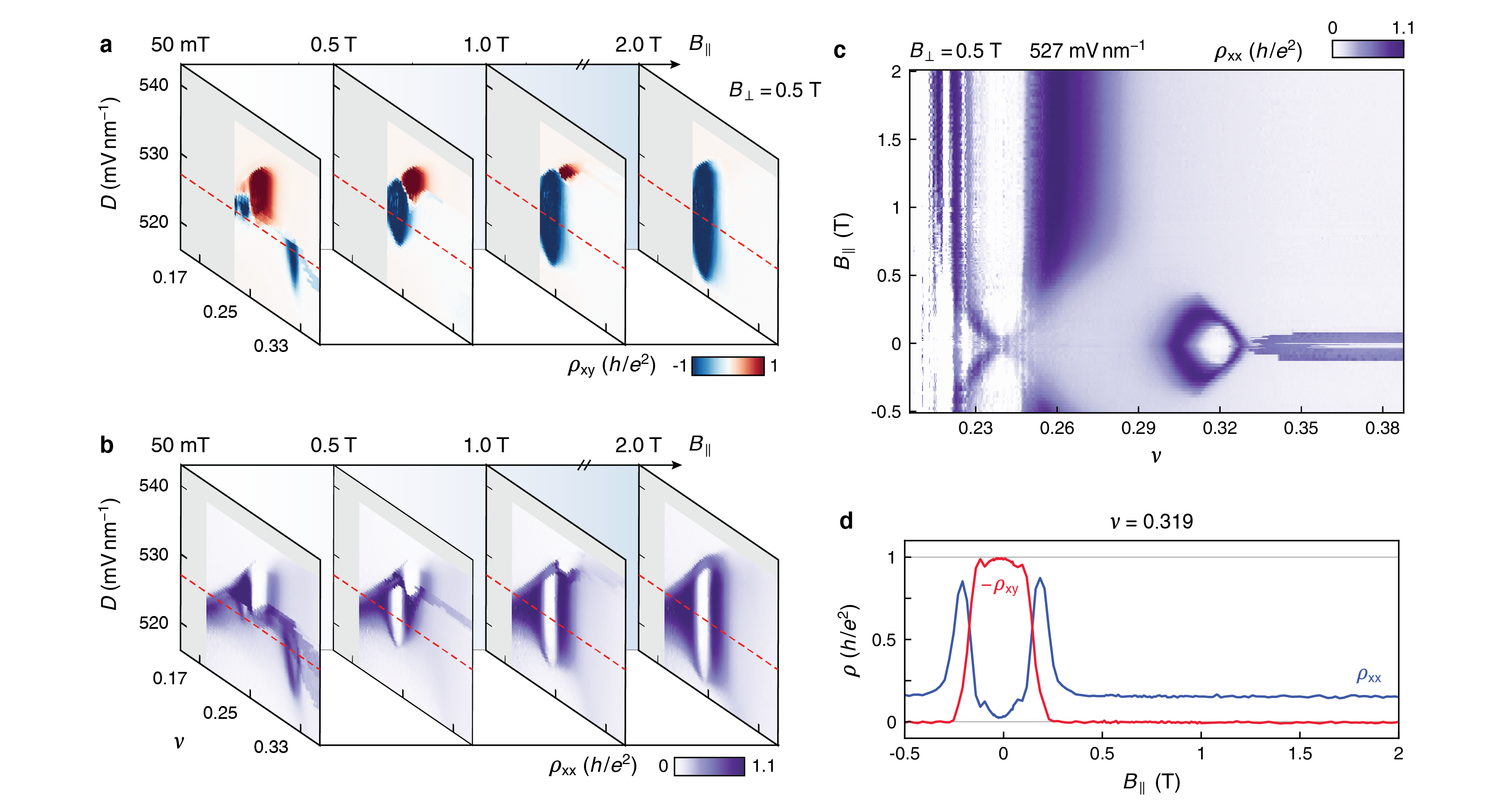}
    \caption{ \textbf{Effect of an in-plane magnetic field on the states at $\nu = 1/4$ and $1/3$.} \textbf{a}, \textbf{b}, $\rho_{xx}$ and $\rho_{xy}$ maps, respectively, obtained at several values of $\bpar$ with fixed $\bperp=0.5$~T, $T = 10$~mK. \textbf{c}, Maps of $\rho_{xx}$ as a function of $\nu$ and $\bpar$ measured at $D = 527$ mV nm$^{-1}$ (denoted by the red dashed lines in \textbf{a}). An in-plane magnetic field stabilizes the \mfour state, whereas the \mthree state is suppressed. \textbf{d}, Measurements of $-\rho_{xy}$ and $\rho_{xx}$ versus $\bpar$ at fixed $\nu = 0.319$, obtained under the same conditions as \textbf{c}. Measurements are obtained at $T = 10$~mK, without (anti)symmetrization.}
    \label{efig:ED_effect_of_Bpar}
\end{figure}

\newpage
\clearpage
\begin{figure}[ht!]
    \centering
    \includegraphics[width = \textwidth]{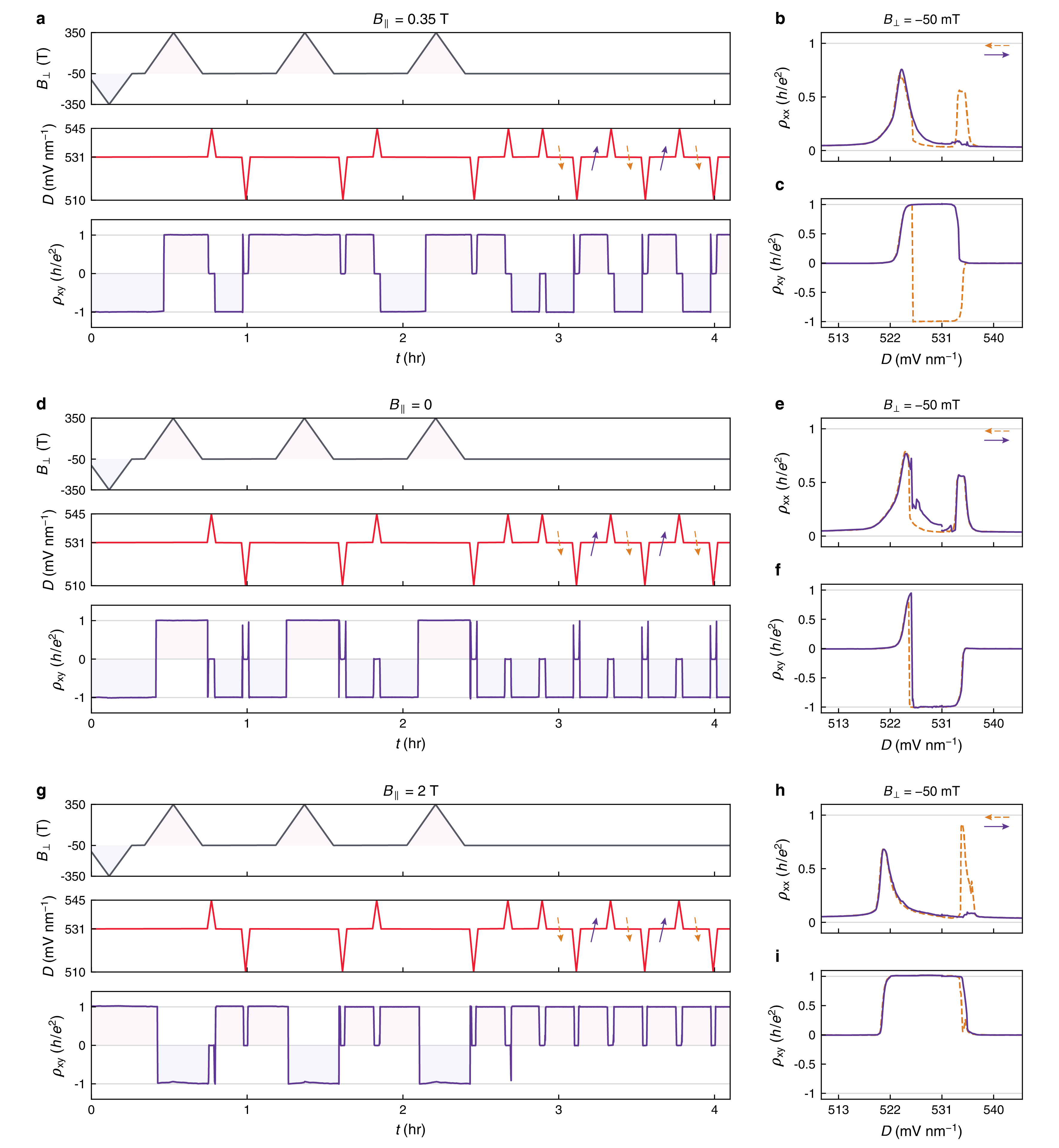}
    \caption{ \textbf{Time-series response of the Hall resistance with $B_{\perp}$ and $D$ sweeps.} 
\textbf{a}, $\rho_{xy}$ (bottom) recorded at $\nu = 0.251$ in response to a sequence of forward and reverse sweeps of $B_{\perp}$ (top) and $D$ (middle), with a fixed $B_{\parallel} = 0.35$~T. The sign of $\rho_{xy}$ switches abruptly to align with the sign of $B_{\perp}$, consistent with a $C = +1$ state. Excursions of $D$ about $531$~mV\,nm$^{-1}$ can also switch the sign of $\rho_{xy}$ (obtained with fixed $B_{\perp} = -50$~mT). \textbf{b}, \textbf{c}, Representative traces of $\rho_{xx}$ and $\rho_{xy}$ measured as $D$ is swept back and forth with $\bperp=-50$~mT. \textbf{d}-\textbf{f}, Analogous measurement to panels \textbf{a}-\textbf{c}, acquired with $B_{\parallel}=0$. In this case, excursions of $D$ about $531$~mV\,nm$^{-1}$ do not switch the sign of $\rho_{xy}$. \textbf{g}-\textbf{i}, Analogous measurements with $B_{\parallel} = 2$~T. In contrast to \textbf{a}-\textbf{f}, the sign of $\rho_{xy}$ in response to $B_{\perp}$ indicates a $C = -1$ state. Excursions of $D$ about $531$~mV\,nm$^{-1}$ also do not change the sign of $\rho_{xy}$. All data was obtained at $T = 300$~mK.
    }
    \label{efig:ED_switching}
\end{figure}

\newpage
\clearpage
\begin{figure*}[!ht]\centering\includegraphics{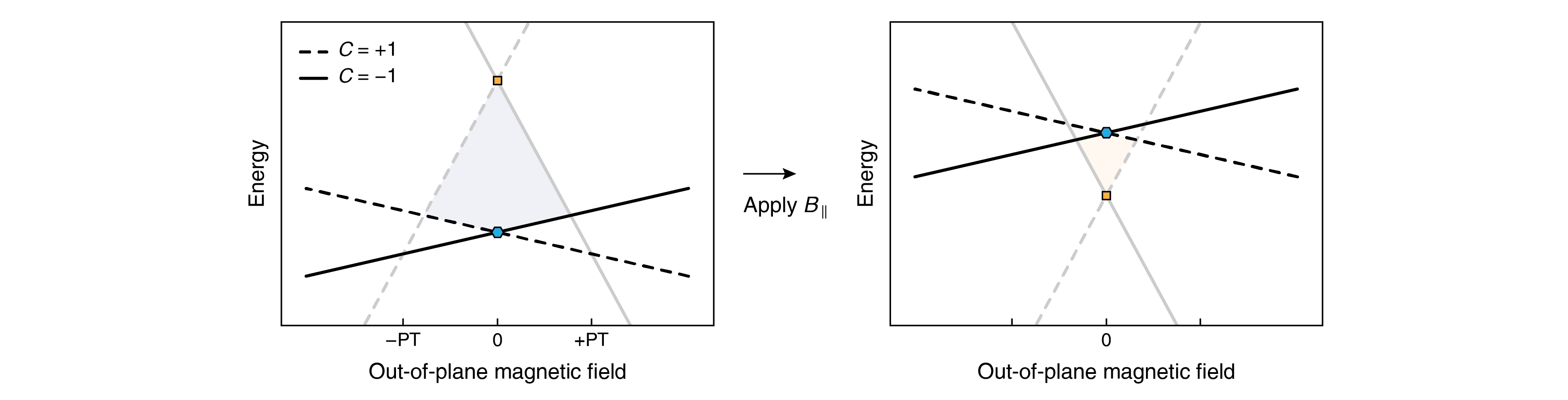}
    \caption{\textbf{Possible two-state model explaining the Chern number reversal of the TEC state at $\nu = 1/4$.}
    Schematic diagrams showing a possible mechanism for the reversal in the sign of $C$ across $\bpt$, as analyzed in Fig.~\ref{fig:2} of the main text, as well as its dependence on $\bpar$ and $D$ as analyzed in Fig.~\ref{fig:4}. The diagram on the left posits the existence of two closely competing TEC states at $\bperp=\bpar=0$, denoted by the blue and yellow markers. The blue circle represents the ground state since it has the lowest energy.\\
    \\
    The model is agnostic to the precise details of the two different states, but assumes that the magnetization of the state marked by the yellow square is larger than that of the state marked by the blue dot, and also that the direction of magnetization for a given sign of Chern number is opposite for the two states. The second assumption is plausible because there is, in general, no fixed relationship between the sign of $C$ and the sign of the total magnetization for orbital magnetic states.  The energies of the $C=+1$ and $-1$ branches of each state evolve oppositely in an applied out-of-plane field, and their magnetization is proportional to the slope of these lines. Above a certain value of $\bperp$, one branch of the state with larger magnetization (yellow square) has a lower energy than either branch of the original ground state (blue circle). The crossing points are denoted as ``$\pm$PT'', indicating the value of $\bperp$ at which a first-order phase transition between the two states is anticipated. For each state, the sign of the Chern number of the branch with lower energy in an applied $\bperp$ corresponds to which has its total magnetization aligned with the external field.\\
    \\
    These are a minimal set of assumptions needed to explain the reversal in the sign of the Chern number across $\bpt$. In this model, the additional reversal with $\bpar$ could result from its coupling to the energetic hierarchy of the two states at fixed $\bperp$. For instance, $\bpar$ could lower the energy of the state marked by the yellow square to a value lower than that of the state marked by the blue circle, thus making the former the ground state. A similar ground state reversal may also result from changing $D$, explaining the gate-induced hysteresis at fixed $\bperp$ and $\bpar$. Although these models are consistent with all of our measurements, we cannot rule out alternative explanations not considered here.
    }
    \label{efig:ED_diagram}
\end{figure*}

\newpage
\clearpage
\begin{figure*}[!ht]
    \centering\includegraphics{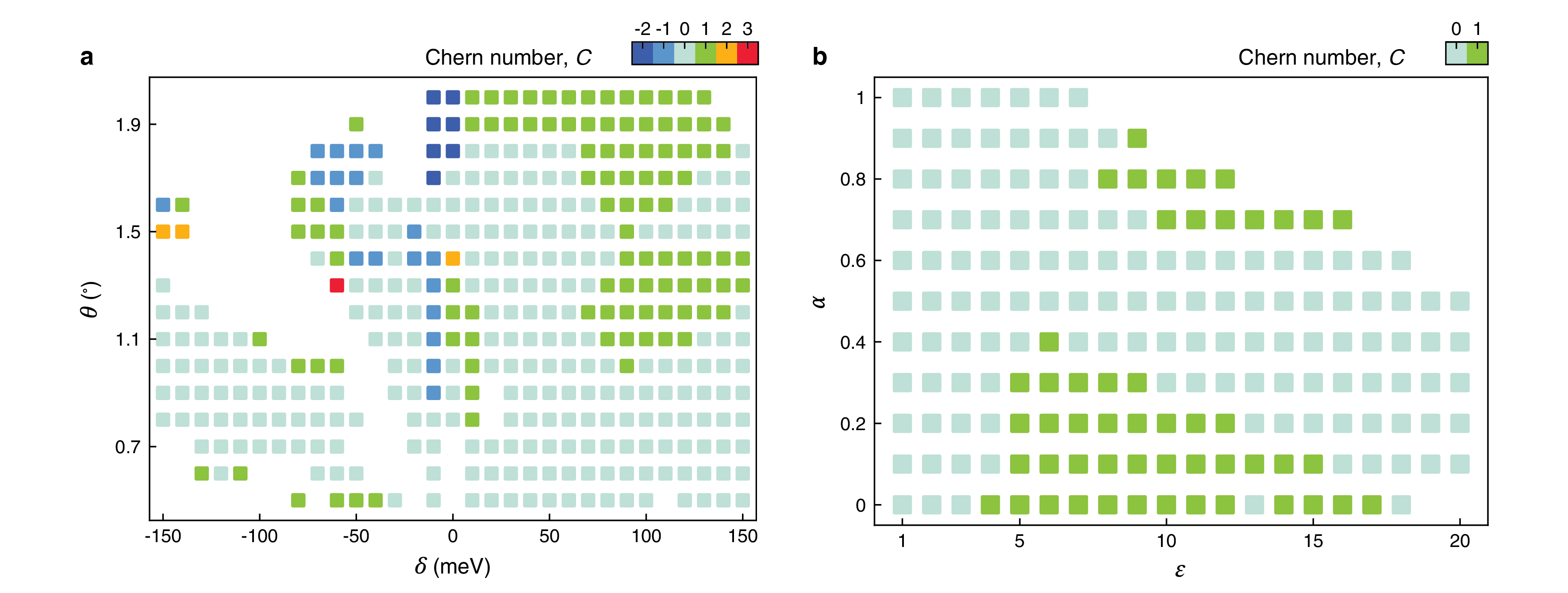}
    \caption{\textbf{Hartree--Fock calculations of the Chern number of the TEC state at $\nu = 1/4$.}
    \textbf{a}, Calculated Chern number as a function of twist angle, $\theta$, and potential difference, $\delta$, with fixed tunneling parameter, $\alpha=0.3$, and dielectric constant, $\epsilon=6$ (see Methods for definitions of parameters). Parameters that do not yield a gap at $\nu = 1/4$ are uncolored. At the twist angle appropriate for our sample, $\theta=1.50^{\circ}$, the model predicts a $C = 1$ state for $\delta=90$~meV. We do not attempt to convert $\delta$ directly to the experimental electric displacement field, $D$, as this would require knowledge of the self-consistent electrostatics of the five graphene layers. We note that to a first approximation, the $\delta$ between layers depends on the applied displacement field $D$ and the spacing between layers, $\delta \approx 4 e D d_{\mathrm{layer}}$, where $4d_{\mathrm{layer}}$ is the spacing between the top of the bilayer and bottom of the trilayer. However, this conversion between $\delta$ and $D$ is not to be taken to be precise: it will be modified by any charges that build up on different layers, which we know to be a significant effect. \textbf{b}, Calculated Chern number as a function of $\alpha$ and $\epsilon$ with fixed $\theta=1.50^{\circ}$ and $\delta=90$~meV.
    }
    \label{efig:ED_Chern_Phase_diagram}
\end{figure*}

\end{document}